\documentclass[nofootinbib]{revtex4}
\usepackage{amsfonts}
\usepackage{amsmath}
\usepackage{amssymb}
\usepackage{mathrsfs}
\usepackage{hyperref}
\usepackage{graphicx}
\usepackage{siunitx}

\usepackage{booktabs}   % <-- booktabs paketini mutlaka yükleyin
\usepackage{tabularx}   % <-- tabularx sütun genişlikleri için
\usepackage{caption}

\usepackage{float}
\usepackage{placeins}
\usepackage{booktabs, tabularx, caption}

\usepackage[dvipsnames]{xcolor}
\providecommand{\U}[1]{\protect\rule{.1in}{.1in}}
\begin{document}
\title{Thermodynamics of Einstein-Euler-Heisenberg Black Holes with Thermal Fluctuations and Nonlinear Electromagnetic Fields}

\author{H. Gürsel \footnote{Corresponding author}}
\email{huriye.gursel@emu.edu.tr}

\author{M. Mangut}
\email{mert.mangut@emu.edu.tr}

\author{E. Sucu}
\email{erdemsc07@gmail.com}

\affiliation{Department of Physics, Eastern Mediterranean
University, Famagusta, 99628 North Cyprus via Mersin 10, Turkey}

\begin{abstract}
This work mainly focuses on the nonlinear Einstein-Euler-Heisenberg theory and its applications from various aspects. Firstly, thermodynamic variables are analytically determined via Smarr formula for a four-dimensional spherically symmetric Einstein-Euler-Heisenberg black hole by taking the Hawking-Bekenstein entropy as the basis. The results are supported by  graphical illustrations for certain Euler-Heisenberg and electric charge parameters, which are in turn used for making further comments on the stability and possible critical points of the concerned black hole. The thermodynamic analyses are then repeated for two distinct cases in which entropy is subject to a logarithmic and an exponential correction, respectively. Our assessments have shown that statistical quantum fluctuations and nonlinear electrodynamic effects can alter the stability and the thermodynamic properties of black holes. Finally, the one-sided bending angle and the gravitational redshift of light are determined in the vicinity of astronomical structures obeying the nonlinear Einstein-Euler-Heisenberg model and the results obtained are applied to three electrically charged, compact stars.
\end{abstract}

\maketitle

\section{Introduction}

Toward the end of the mid-19th century, the theory of classical electrodynamics was constructed by James Clerk Maxwell \cite{maxwell1865viii}, through which Faraday's experimental discoveries were explained via the unification of electricity and magnetism. Maxwell's equations combined with preceding and subsequent work carried out by numerous scientists eventually resulted in the establishment of classical field theory. It is worth noting that electrodynamic theories in curved spacetime have been extensively examined by various authors, leading to the successful explanation of several novel physical phenomena. Notably, a foundational contribution by Rainich \cite{rainich1925electrodynamics}  explores the role of electrodynamics within the context of general relativity. Furthermore, in the seminal work of Misner and Wheeler on geometrodynamics \cite{misner1957classical}, the Einstein-Maxwell equations are formulated as a fundamental Lagrangian framework in which particles are interpreted as geons. In both studies, black hole solutions are investigated as part of the broader theoretical structure. However, despite all the beauty and predictions offered by the theory, Maxwell's equations seemed to remain insufficient in explaining the Sauter-Schwinger effect  \cite{sauter1931verhalten,schwinger1951gauge} observed almost a century later, which briefly refers to the formation of particle-antiparticle pairs in vacuum in the presence of a strong electric field. This phenomenon can be considered as an extension of Dirac's relativistic quantum  theory of the electron \cite{dirac1928quantum} giving birth to further pioneering developments in both theoretical and experimental physics.\\

In addition to the lack of the theoretical understanding behind the Sauter-Schwinger effect, when the electric field produced by a point electric charge was examined, the self-interaction was found to be not well defined, suggesting that classical electrodynamics required corrections, although recent studies such as \cite{lopez2020electrodynamic,lopez2020stability}  have demonstrated that the issue of infinite self-energy in Maxwell's electrodynamics can be mitigated through the use of extended charge distributions. To overcome the point charge singularity, a nonlinear modification was carried out by Born and Infeld \cite{born1934foundations}. While the implementation of this modification resolved the singularity issue, the Born-Infeld model is still classified as a purely classical theory, since no quantum-mechanical principles are involved. Some studies on the Born-Infeld theory can be found in \cite{born1934foundations,garcia1984type,fradkin1985non,salazar1987duality,gibbons1995electric,deser1998born,ayon1998regular}. The inclusion of quantum mechanics in the theory of electrodynamics took place with the work of Euler and Heisenberg \cite{Heisenberg:1936nmg} approximately two years after the proposal of the Born-Infeld model. {The one-loop effective Lagrangian was first calculated by Euler and Heisenberg \cite{Heisenberg:1936nmg}, and later by Weisskopf \cite{weisskopf1939self}, which was then extended to two loops by Ritus \cite{ritus1975lagrangian}. The nonlinear Euler-Heisenberg Lagrangian is the zero-frequency, infinite-wavelength limit of the general nonlinear effective Lagrangian of quantum electrodynamics (QED), which serves as the generating functional of one-photon-irreducible vertex functions. For a general discussion, see \cite{weinberg1995quantum}.Correspondingly, the vacuum polarisation and photon-photon scattering exposed to strong electromagnetic fields, which are obtained by differentiating Euler-Heisenberg Lagrangian over fields, are  the small-momentum asymptotes of the needed functions. Polarisation tensor in an arbitrary constant field beyond the mass shell of a photon is calculated with one-loop accuracy within the Furry picture in \cite{batalin1971green} and the special case of a magnetic field taken alone is included in \cite{shabad1975photon}. For the same in a plane-wave field, one may refer to \cite{ritus1985quantum}. Although photon-photon scattering exists without the need of an external field, phenomena such as photon-photon merging or photon splitting do require an external field, and its calculation in a magnetic field using Euler-Heisenberg Lagrangian has been carried out by Birulas \cite{bialynicka1971nonlinear}. Furthermore, the significant contribution by S. Adler et al. \cite{adler1970photon}, which established results valid for finite photon momenta, has been widely utilised in understanding astrophysical phenomena in strong electromagnetic fields, including those encountered in neutron stars \cite{mosquera2004non,kim2011light}. It is also worthwhile to note that if the Euler-Heisenberg Lagrangian is truncated at any finite order of its Taylor expansion in powers of the field, the electromagnetic field energy of a point charge converges \cite{costa2015finite}. For general convergence conditions, one may see \cite{adorno2017quantum}.\\

Recent studies have shown interesting applications of Euler-Heisenberg nonlinear electrodynamics (NED) in various fields. However, at this stage, it is natural to consider the relative advantages of the Euler–Heisenberg framework in comparison to the Born–Infeld model in the context of black hole solutions in NED (for instance, see \cite{gullu2021black,gullu2021double}  for a  NED model based on the Born–Infeld framework). The Euler–Heisenberg theory is particularly well suited to the study of quantum corrections to black hole thermodynamics, the propagation of light in strong-field regimes and semiclassical backreaction phenomena. If one's focus lies in exploring statistical fluctuations in regimes where the field strengths are strong but remain below the critical (Schwinger) QED threshold, the Euler–Heisenberg framework, derived from quantum electrodynamics, offers a more physically appropriate and predictive approach than the Born–Infeld model. In \cite{ruffini2013einstein}, Ruffini et.al. formulated Einstein-Euler-Heisenberg theory and inspected the corresponding nonrotating black hole solutions with electric and magnetic charges in spherical geometry by taking into account the Euler-Heisenberg effective Lagrangian of one-loop nonperturbative QED contributions. The QED one loop effective Lagrangian \footnote{Following \cite{batalin1971green,shabad1975photon,ritus1985quantum,adler1970photon}, we note that the Euler–Heisenberg Lagrangian accounts only for the local part of the QED effective action. The full description of QED nonlinearity, especially for light propagation in nonuniform strong fields, necessitates incorporating the nonlocal contributions dependent on spacetime derivatives of the field.}
\begin{equation}
\Delta \mathscr{L}_{\text{eff}} = \frac{1}{8\pi^2} \int_0^{\infty} \frac{ds}{s^3} 
\left[ e^2 \epsilon \xi s^2 \coth(e \epsilon s) \cot(e \xi s) 
- \frac{e^2}{3}(\epsilon^2 - \xi^2)s^2 - 1 \right] 
e^{-i s (m_e^2 - i \eta)}\label{hey}
\end{equation}
was obtained by Euler and Heisenberg for the first time (for further details on the original discussion, one may see \cite{Heisenberg:1936nmg}. Here, $\eta$ stands for  a positive infinitesimal parameter, the scalar $X$ and the pseudoscalar $Y$ are Lorentz invariants which can be expressed in terms of the field strength $F_{\mu\nu}$ and and its dual as
\begin{equation}
\begin{aligned}
X &\equiv -\frac{1}{4} F_{\mu\nu} F^{\mu\nu} = \frac{1}{2}(\mathbf{E}^2 - \mathbf{B}^2), \\
Y &\equiv -\frac{1}{4} F_{\mu\nu} \tilde{F}^{\mu\nu} = \mathbf{E} \cdot \mathbf{B},
\end{aligned}
\end{equation}
One shall note that the two alternative Lorentz invariants $\epsilon$ and $\xi$ via the relations
\begin{equation}
\begin{aligned}
&\epsilon = \sqrt{ \left(X^2 + Y^2\right)^{1/2} + X }, \\
&\beta = \sqrt{ \left(X^2 + Y^2\right)^{1/2}  - X }.
\end{aligned}
\end{equation}
In the cases when the electromagnetic fields of concern obey $\varepsilon/E_{\rm c} \ll 1$ and $\xi/E_{\rm c} \ll 1$, $E_c=m_e^2/e$ representing the critical field in natural units, the real part of the QED one-loop effective Euler-Heisenberg Lagrangian is determined under the weak-field approximation, and as a result, the Einstein-Euler-Heisenberg action can be expressed in terms of the Lorentz invariants as
\begin{equation}
\mathcal{I}_{\rm EEH}
= -\frac{1}{16\pi} \int d^4x\, \sqrt{-g}\, R
\;+\; \int d^4x\, \sqrt{-g}
\left[
X
+ \frac{2\alpha^2}{45 m_e^4} \left(4X^2+7P^2\right)
\right],\label{bes}
\end{equation}
to the leading order in $\alpha^2$. Here, $R$ represents the Ricci scalar, $\alpha$ is the fine structure constant and $\mu_{\text{one-loop}}=\frac{\alpha}{45\pi E_c^2}$ is derived from the one-loop effective Lagrangian whose details will be presented in the upcoming section. Consequently, the sextic term appearing in the Einstein-Euler-Heisenberg metric function derived via the Einstein field equations will be expressed in an explicit form. Our analyses will then show that, under the introduction of a rescaled dimensionless parameter $\mu$ via $\mu_{\text{one-loop}}\rightarrow \mu \equiv \bar{\mu}/M^2$ with $\bar{\mu}=\kappa \mu_{\text{one-loop}}$ and $\kappa$ a dimensionless scale factor, a thorough inspection in parameter space can be achieved. Subsequently, this dimensionless parameter will be shown to reveal physically reasonable blackhole structures, once it falls within the range $0 \leq \mu \leq 50/81$. Yet, throughout the analysis of black hole thermodynamics via the Smarr formula, one needs to consider the dimensional $\bar{\mu}$, provided that its impact on black hole thermodynamics is desired to be addressed. Although ${\mu}_{\text{one-loop}}$ is derived from the weak-field approximation by considering the leading order to be up to $\alpha^2$, one shall note that this can be subject to changes once conditions differ. For a similar approach, one is recommended to see \cite{magos2020thermodynamics}.\\

In \cite{breton2021birefringence}, birefringence and quasinormal modes of Einstein-Euler-Heisenberg black holes are studied in the eikonal approximation; whereas in \cite{magos2020thermodynamics}, the thermodynamic properties and phase transitions are examined with the inclusion of cosmological constant in the Euler-Heisenberg theory. Furthermore, a NED theory whose action involves Maxwell, weak Euler-Heisenberg and strong electromagnetic terms is analysed for the electrostatic case with the aim of investigating the corrections of Coulomb's law of a point charge in \cite{bermudez2020coulomb}. The work of Faber in which  divergencies are addressed  via soliton theory \cite{faber2012particles} is crucial in this respect, as well as the seminal papers of  Carter \cite{carter1968global} and Burinskii \cite{burinskii2012gravity,burinskii2008dirac}. Some brief discussion concerning the possibility of modified Kerr- Newman black hole solutions to represent fundamental particles can be found in \cite{carter1968global,burinskii2012gravity}. Also, numerical methods (one of which can be found in \cite{lindner2023numerical}) have been constructed to solve NED equations of Euler-Heisenberg weak field expansion, and consequently, simulations of nonlinear quantum vacuum are established. In another study \cite{krug}, detailed analyses are carried out on charged black hole solutions in Euler-Heisenberg type NED with two parameters.\\

Black hole thermodynamics is considered as a prominent theoretical framework that combines laws of thermodynamics with principles of general relativity and quantum mechanics\cite{york1986black,myers1988black,dolan2011pressure,anabalon2018holographic}. In addition, the interconnection between the thermodynamic concepts and the geometrical properties of black holes play a vital role in the pursuit of developing a quantum theory of gravity. The study of concepts such as Bekenstein-Hawking entropy \cite{bekenstein2020black,bekenstein1973black,hawking1975particle} and Hawking radiation \cite{hawking1974black} raises intriguing questions about the information content and the thermodynamic future of the universe. Some examples of such discussions can be found in \cite{hawking1976breakdown,penrose1979singularities}. It is also worthwhile noting that the merging of thermodynamics and general relativity enables one to carry out the stability analysis of solutions of Einstein's equations by examining the potentials of the concerned thermal systems, as well as the evaluating thermodynamic parameters such as the heat capacity ($C$), a critical term in the determination of the overall stability (negative $C$ indicating thermodynamic instability \cite{hawking1983thermodynamics}), the enthalpy ($H$), the internal ($E$), Helmholtz ($F$) and Gibbs ($G$) free energies\cite{greiner2012thermodynamics}. One can also use the stability analysis to assess phase transitions, since the divergence of $C$ is usually an indicator of a critical point.\\ 

To evaluate the thermodynamic parameters, one would need to have a convenient expression for the entropy of the black hole structure of concern. With the development of principles of AdS/CFT correspondence, loop quantum gravity, string theory, and other frameworks, the Bekenstein-Hawking entropy has been receiving various modifications over the years. For some examples, one can refer to \cite{strominger1996microscopic,das2002general,alsing2018massive,mahapatra2011black,pourhassan2024non}. By combining the discussions of statistical mechanics with the models mentioned above, the corrected entropy of a black hole can be written as \cite{dehghani2021quantum}
\begin{equation}
S=S_0+\xi ln f_1(S_0)+\frac{\gamma}{S_0}+\eta e^{-S_0
}, \label{S}
\end{equation}
where $S_0$ is the Bekenstein-Hawking entropy such that $S_0=\mathcal{A}/4l^2_P$. Here, $\mathcal{A}$ refers to the horizon area and $l_P$ is the Planck length. The logarithmic and the inverse area terms arise due to small stable fluctuations around equilibrium \cite{das2002general}, whereas the exponential contribution is treated as a non-perturbative quantum correction \cite{dabholkar2015nonperturbative}. Although there exist a large number of studies in the literature on the derivation and implications of the logarithmic and inverse area terms\cite{solodukhin1998entropy,Kastrup:1997iu,mann1998universality,medved1999quantum,kaul2000logarithmic,medved2001one,gour2003thermal,chatterjee2004universal,rovelli1996black,domagala2004black,medved2004comment,berti2005estimating,medved2005hawking,zhang2008black,iliesiu2022revisiting} , the exponential correction has less thoroughly been investigated (see\cite{chatterjee2020exponential,pourhassan2024non} for some examples).\\

In this study, the thermodynamic potential of an Einstein-Euler-Heisenberg black hole will be determined by considering the logarithmic and exponential corrections mentioned in $(\ref{S})$ separately. Doing so allows one to examine two distinct black hole states, the small- and large-horizon regimes, independently. Then, for each case, the relevant thermodynamic parameters will be analytically evaluated and stability analyses will be carried out accordinly. For small horizons, the horizon area obeys $A\sim  l_p^2$, which in turn makes the corrected entropy  reduce to \cite{chatterjee2020exponential}
\begin{equation}
S\sim  S_0+\eta e^{-S_0}\label{ent}.
\end{equation}

In \cite{chatterjee2020exponential}, Chatterjee and Ghosh derived a general equation for the entropy of any black hole  with an isolated horizon in the case where the horizon area is small. Rather than using the tools of string theory and loop quantum gravity, the authors based their discussions on the horizon geometry. The entropy derived with the concerned method is consistent with $(\ref{ent})$.\\

In contrast, if the horizon area is relatively large, the entropy becomes \cite{gour2003thermal}

\begin{equation}
S\sim S_0+\xi ln f_1(S_0). \label{SSS}
\end{equation}

 An entropy corrected in this form is considered for charged anti-de Sitter \cite{pourhassan2015thermal}, Horava-Lifshitz \cite{pourhassan2018quantum}, singly spinning Kerr-AdS \cite{pourhassan2016thermodynamics} and symmergent  \cite{ali2023thermodynamics}  black holes. Furthermore, the effect of these small fluctuations around equilibrium on thermodynamic parameters is studied for $\mathcal{F}(R,\mathcal{G})$ gravity black holes with constant topological Euler density in NED \cite{mangut2024thermal}, Kerr-Newman-AdS and Reissner-Nordström-AdS black holes \cite{zhang2018corrected}.\\

Gravitational lensing, a strong observational tool in testing the validity of models constructed based on Einstein's theory of general relativity, can also be used to assess models emerging out of NED\cite{de2023electrically,javed2020effect,fu2021weak,gurtug2019effect}. In the majority of the cases, the mathematical complexity of NED equations yields black hole solutions including non-trivial functions, which may eventually give rise to difficulties when it comes to determining the associated lensing angle theoretically. Such challenges can be handled by using the generalisation of the inner product to curved spacetimes proposed by Rindler and Ishak \cite{rindler2007contribution}. This method has been shown to cause nonlinear modifications to the purely electrical solutions of Einstein-Power-Maxwell NED theory, as well as to the purely electrical and purely magnetic solutions of the non-linear Kruglov model \cite{gurtug2019effect,gurtug2020gravitational}.\\ 

One of the most significant applications of the Rindler-Ishak method is the determination of the gravitational lensing angle around compact astronomical objects. In this paper, the gravitational lensing angles in the vicinity of VelaX-1, SAXJ1808.4-3658 and 4U1820.30 will be evaluated. During our calculations, the numerical values of the mass, charge and radius of the concerned objects will be taken from \cite{ilyas2018charged}. Since all of the objects specified above include strong electromagnetic fields, the metric of choice will be determined based on the Euler-Heisenberg NED theory, which will enable us to use our theoretical results in observational applications. The method of Rindler and Ishak is generalised in \cite{gurtug2021gravitational} for rotating metrics and its application showed indications of cosmic voids in rotating Bertotti-Robinson spacetimes. Moreover, the effect of Lorentz symmetry breaking on the  lensing angle is analytically analysed for rotating Bumblebee black holes \cite{mangut2023probing} and the same methodology is used to calculate the lensing angle in the vicinity of a Kerr-Newman-AdS black hole, together with some applications \cite{mangut2023gravitational}.\\ 

This study consists of five sections. In Section \ref{sec1}, the physical properties of the Einstein-Euler-Heisenberg black hole will be provided briefly, which will then be followed by a detailed analysis of its thermodynamic characteristics constituting Section \ref{sec3}. The analytical solutions obtained will be supported by graphical illustrations for further analysis. Section \ref{sec4} will focuse on the astrophysical applications of the Euler-Heisenberg NED theory by examining gravitational lensing and gravitational redshift. Finally, conclusive remarks will be presented in the last section.

\section{Einstein-Euler-Heisenberg Black Hole Solution}\label{sec1}

Having addressed the QED one-loop Euler-Heisenberg effective Lagrangian and its constituents, Ruffini et.al. \cite{ruffini2013einstein} introduced the effective Lagrangian of nonlinear EM fields as
\begin{equation}
\mathscr{L}_{\text{eff}} = \mathscr{L}_{M} + \Delta \mathscr{L}_{\text{eff}},\label{hey1} 
\end{equation}
where  $\mathscr{L}_M$ is the Maxwell Lagrangian and  $\Delta \mathscr{L}_{\text{eff}}$ is as given in \eqref{hey}, whose real part reads 
\begin{equation}
(\Delta\mathcal{L}^{\cos}_{\text{eff}})_\mathcal{P} = \frac{1}{8\pi^2} \sum_{n,m=-\infty}^{\infty} \frac{1}{\tau_m^2 + \tau_n^2} \left[ \bar{\delta}_{m0} J(i\tau_ m m_e^2) - \bar{\delta}_{n0} J(\tau_n m_e^2) \right],
\end{equation}
in which $\tau_n=n\pi/e\epsilon$ and $\tau_m=m\pi/e\epsilon$. The symbol $\bar{\delta}_{ij} = 1 - \delta_{ij}$ represents the complimentary Kronecker delta and  
\begin{equation}
J(z) \equiv \mathcal{P} \int_0^{\infty} ds \frac{s e^{-s}}{s^2 - z^2} = -\frac{1}{2} \left[ e^{-z} \mathrm{Ei}(z) + e^z \mathrm{Ei}(-z) \right].
\end{equation}
Here $\mathcal{P}$ stands for the principle value integral, whereas $\mathrm{Ei}(z)$ is the exponential-integral function 
\begin{equation}
\mathrm{Ei}(z) \equiv \mathcal{P} \int_{-\infty}^{z} \frac{e^{t}}{t} \, dt = \ln(-z) + \sum_{k=1}^{\infty} \frac{z^k}{kk!}.
\end{equation}
In the weak EM field limit where $\epsilon/E_c<<1$ and $\xi/E_c<<1$ are satisfied, the weak-field expansion of the one loop effective Lagrangian becomes
\begin{equation}
(\Delta \mathcal{L}^{\cos}_{\text{eff}})_\mathcal{P} = \frac{2\alpha^2}{45 m_e^4} \left(4 X^2 + 7 Y^2\right) +\mathcal{O}\left( \alpha^3\right).
\end{equation}
Under such circumstances, the effective Lagrangian of nonlinear EM fields presented \eqref{hey1} takes the form  
\begin{equation}
\mathcal{L}_{\text{EH}}= X+(\Delta \mathcal{L}^{\cos}_{\text{eff}})_\mathcal{P},\label{hey2}
\end{equation}
in the weak-field approximation $\left(\mathcal{L}_{\text{EH}} = \left. \mathcal{L}_{\text{eff}} \right|_{\text{large } z}\right)$ which in turn gives rise to the Einstein-Euler-Heisenberg action 
\begin{equation}
\mathcal{I}_{EEH} = -\frac{1}{16\pi G} \int d^4 x \, \sqrt{-g} R+ \int d^4 x \, \sqrt{-g} \mathcal{L}_{\text{EH}} , 
\label{izm1}
\end{equation}
where $\mathcal{L}_{\text{EH}}$ is given in \eqref{hey2}. To obtain black hole solutions of the concerned theory, one shall analyse Einstein field equations
\begin{equation}
G^{\mu\nu}=R^{\mu\nu}-\frac{1}{2}g^{\mu\nu}R=8\pi GT^{\mu\nu},\label{ein}
\end{equation}
whose energy-momentum tensor reads
\begin{equation}
T^{\mu\nu}=\frac{2}{\sqrt{-g} }\frac{\delta \mathcal{I}_{EEH} }{\delta g_{\mu\nu}}.
\end{equation}
For electromagnetic fields varying smoothly over space, the energy-momentum tensor becomes
\begin{equation}
T^{\mu\nu}=T^{\mu\nu}_M\left(1+\mathcal{A}_X \right)+g^{\mu\nu}\left[\mathcal{A}_X X+\mathcal{A}_Y Y- (\mathcal{L}^{\cos}_{\text{eff}})_\mathcal{P} \right], \label{hey6}
\end{equation}
where $T^{\mu\nu}_M=-g^{\mu\nu}X+F^{\mu}_{\lambda}F^{\lambda\mu}$ represents the energy-momentum tensor of the electromagnetic fields of linear Maxwell theory and two invariants $\mathcal{A}_X$ and $\mathcal{A}_Y$ are defined as
 \begin{equation}
\mathcal{A}_X=\frac{\delta\left(\mathcal{L}^{\cos}_{\text{eff}}\right)_{\mathcal{P}} }{\delta X} \;\; \text{and} \;\; \mathcal{A}_Y=\frac{\delta\left(\mathcal{L}^{\cos}_{\text{eff}}\right)_{\mathcal{P}} }{\delta Y} .
\end{equation}
In this work, we are interested in case of vanishing magnetic field which indicates that the conditions, $\xi=Y=0$, $\epsilon=E$, $X=E^2$ and $\mathcal{A}_Y=0$ shall be satisfied. Therefore, the weak-field effective Lagrangian  \cite{ruffini2013einstein} reduces to
 \begin{equation}
\left(\Delta\mathcal{L}^{\cos}_{\text{eff}}\right)_{\mathcal{P}}= \frac{2\alpha^2}{45m_e^4}(4X^2)+\mathcal{O}\left( \alpha^3\right) \label{hey3}
\end{equation}
substituting \eqref{hey3} in the definition of $\mathcal{A}_X$ and $\mathcal{A}_Y$ one obtains
\begin{equation}
\begin{aligned}
\mathcal{A}_X &\approx  \frac{2\alpha^2}{45m_e^4}(8X) \\
\mathcal{A}_Y &\approx \frac{2\alpha^2}{45m_e^4}(14Y) \label{hey4}
\end{aligned}
\end{equation}
up to $\mathcal{O}\left( \alpha^3\right)$ in the weak-field approximation. As we are interested in the $B=0$ case, these expression further reduce to
\begin{equation}
\begin{aligned}
\mathcal{A}_X &\approx  \frac{2\alpha^2}{45m_e^4}(8X) \\
\mathcal{A}_Y &= 0. \label{hey5}
\end{aligned}
\end{equation}
Once \eqref{hey5} is plugged into Eq.\eqref{hey6}, the energy-momentum tensor becomes
 \begin{equation}
T^{\mu\nu}= T^{\mu\nu}_M \left[1+8\left(  \frac{2\alpha^2}{45m_e^4}X\right) \right]+g^{\mu\nu}\left[  \left(\frac{2\alpha^2}{45m_e^4}\right)4X^2\right]+\mathcal{O}\left( \alpha^3\right).
\end{equation}
Recalling that $X=E^2/2$ for the $B=0$ case, one can also write
 \begin{equation}
T^{\mu\nu}= T^{\mu\nu}_M \left[1+4\left(  \frac{2\alpha e^2E^2}{45\pi m_e^4}\right) \right]+g^{\mu\nu}\left[  \left(\frac{\alpha e^2 E^4}{90\pi m_e^4}\right)\right]+\mathcal{O}\left( \alpha^3\right),
\end{equation}
To investigate a spherically symmetric black hole solution one shall introduce
\begin{equation}
ds^{2}=f(r)dt^{2}-\frac{dr^{2}}{f(r)}-r^{2}\left(d\theta ^{2}+sin^{2}(\theta)d\phi ^{2}\right),
\label{metric1}
\end{equation}
such that $f(r)\equiv1-2\frac{m(r)}{r}$ in natural units. In such a case, the gauge potential reads $A_{\mu}=\left[A_0(r),0,0,0 \right]$ leading to an electric field in the form $E(r)=-\partial A_0(r)/\partial r$. Both $f(r)$ and $E(r)$ fulfill the Einstein and the electromagnetic field equations. Hence, going back to Einstein equations \eqref{ein},one can write
\begin{equation}
\frac{f(r)}{r^2}(1-f(r)-rf'(r))=8\pi T^{00}, \label{rotti}
\end{equation}
with $T^{00}$ representing the local energy density of the electromagnetic field. To obtain the metric function explicitly, the $tt$-component of Einstein equations is required to be checked. For a static spherically symmetrical metric, the mass function can be computed via a detailed analysis of
\begin{equation}
G^{00}=\frac{f(r)}{r^2}(1-f(r)-rf'(r)). \label{rottihey}
\end{equation}
Upon performing the necessary algebraic manipulations for $B=0$ ($T^{00}_M= E^2/2$), one obtains
\begin{equation}
m(r)\approx M-\frac{\bar{Q}^2}{8\pi r}\left[1-\frac{\alpha \bar{Q}^2}{3600\pi^3E_c^2r^4}\right]. \label{rottiss}
\end{equation} 
where $\bar{Q}=\sqrt{4\pi}Q$ for consistency with studies such as \cite{magos2020thermodynamics,breton2021birefringence,luo2022gravitational}.This expression plays a significant role in the analysis of QED effects on the black hole solutions. Ultimately, from $f(r)=1-2m(r)$, the metric function becomes
\begin{equation}
f(r)=1-\frac{2M}{r}+\frac{Q^2}{r^2}\left\{1- \frac{\bar{\mu}Q^2}{20r^4}\right\}. \label{nmetric}
\end{equation}
Note that under this construction, the dimensionful parameter $\bar{\mu}=\kappa\mu_{\text{one-loop}}$ can be treated as the NED parameter of the concerned framework.\\

To find the real positive roots, one can set $f(r)\big\vert_{r=r_H} = 0$ which leads to 
\begin{equation}
r_H^6 - 2Mr_H^5 + Q^2 r_H^4 - \frac{\bar{\mu}Q^4}{20} = 0, \label{root}
\end{equation}
where $r_H$ represents the black hole horizon and the nonlinear Euler-Heisenberg parameter has dimension $[\bar{\mu}]=L^2$. For simplicity, the rescalings $M \to M/M$, $r_H \to  r_h \equiv r_H/M$, $Q \to q \equiv  \; Q/M$ and $\bar{\mu} \to \mu \equiv  \; \bar{\mu} /M^2 $ are applied, enabling one to get rid of the mass parameter and express Eq.\eqref{root} in terms of the dimensionless quantities as follows.
\begin{equation}
r_h^6 - 2r_h^5 + q^2 r_h^4 - \frac{\mu q^4}{20} = 0. \label{root2}
\end{equation}
At this point, let us name the sixth-degree polynomial as
\begin{equation}
f(r_h)=r_h^6 - 2r_h^5 + q^2 r_h^4 - \frac{\mu q^4}{20}, \label{root3}
\end{equation}
and evaluate
\begin{equation}
\frac{df(r_h)}{dr_h}=0, \label{root4}
\end{equation}
which results in
\begin{equation}
6r_h^5 - 10r_h^4 + 4 q^2 r_h^3=0. \label{root5}
\end{equation}
Taking the common factor out, one can write
\begin{equation}
2r_h^3(3r_h^2 - 5r_h + 2 q^2)=0. \label{root6}
\end{equation}
To ensure physical consistency, one shall focus on the non-zero solutions which can be obtained via solving
\begin{equation}
3r_h^2 - 5r_h + 2 q^2=0. \label{root7}
\end{equation}
Consequently, the physically meaningful horizon range is found to be within
\begin{equation}
\underbrace{\frac{5-\sqrt{25-24 q^2}}{6}}_{\text{$r_-$}}\leq r_h \leq \underbrace{\frac{5-\sqrt{25-24 q^2}}{6}}_{\text{$r_+$}}. \label{root8}
\end{equation}
leading to the condition $q^2\leq25/24$. Given that $q$ corresponds to the charge to mass ratio of the concerned black hole, realistically, the upper limit of this ratio is estimated as $\sqrt{25/24}$.\\

Having determined the physical upper bound of the charge parameter, let us now figure out whether any constraint exists for the Euler-Heisenberg parameter by investigating the limiting cases of Eq.\eqref{root3} where horizons coincide and become degenerate. At this point, it is worthwhile to note that beyond these thresholds, one would get a naked singularity rather than a black hole solution. Setting $f\left(r_h=r_{\pm} \right)=0$, one can write
\begin{equation}
 \mu_{\pm}=\frac{5^5\left(1\pm\sqrt{1-\frac{24}{25}q^2}\right)^4\left(6q^2-5\left(1\mp\sqrt{1-\frac{24}{25}q^2}\right)\right)}{18^3q^4} \label{root9}
\end{equation}
When the charge parameter is taken as $ q=\sqrt{25/24}$, $\mu_+$ and $ \mu_-$ coincide. Under this choice, the Euler-Heisenberg parameter reaches its physically meaningful maximum which reads
\begin{equation}
  \mu_{max}=\frac{50}{81}.
\end{equation}
For similar analyses based on different black hole structures, one can check \cite{breton2021birefringence,toshmatov2017rotating}.\\

In brief, for a black hole solution to exist, $ q^2\leq25/24$ and $\mu\leq50/81$ are both required to be satisfied. These analytically evaluated constraints can also be determined graphically, once Eq.\eqref{root3} is tackled numerically. According to Descartes' rule of signs and the nature of $f(r_h)$, the solutions allow from zero to three horizons, depending on the charge and Euler-Heisenberg parameter values.

\begin{figure}[H]
\centering
\includegraphics[width=7cm]{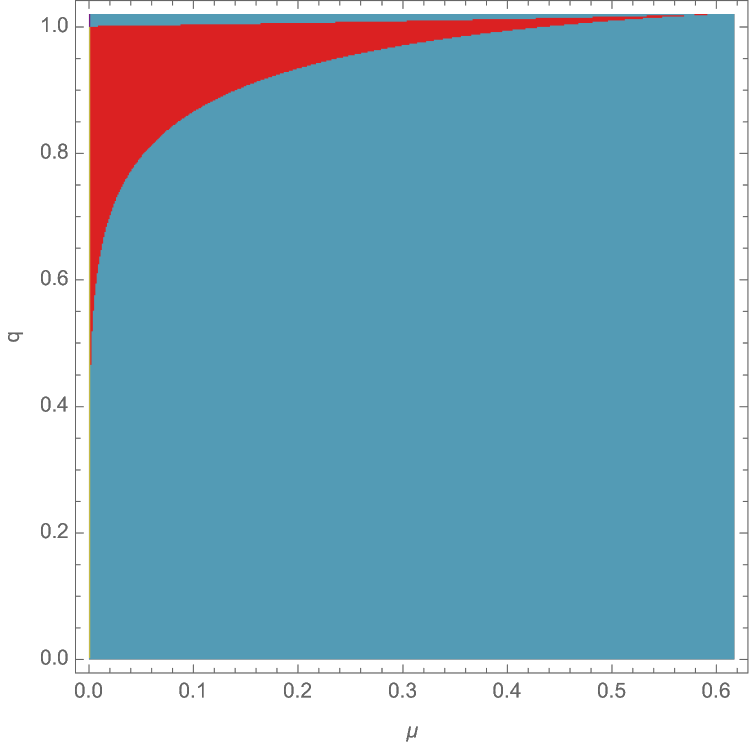}  
\caption{This figure displays possible black hole solutions in parameter space $(\mu,q)$. The red regions correspond to black holes with three horizons, whereas those in blue represent conditions for naked singularity. Once the Euler-Heisenberg parameter is set to zero, there exist certain charge parameter values at which one or two horizons are supported (yellow-green regions) which can not be seen due to the scale.}
\label{hi}
\end{figure}

To further clarify the role of the NED correction associated with the Euler-Heisenberg term for $q^2\leq25/24$ and $\mu\leq50/81$, the structure of the metric function (\ref{nmetric}) is examined. It is important to note here that the root-finding analysis has been conducted by dimensionless reparametrisation of the relevant parameters. In this context, when reverting to dimensional parameters, the corresponding maximum limit boundaries become $Q^2/M^2\leq25/24$ and $\kappa \mu_{\text{one-loop}}/M^2\leq50/81$, respectively. In this work, although the Euler–Heisenberg parameter $\bar{\mu}_{\text{one-loop}}$ is a small and fixed QED vacuum polarisation correction, we reinterpret it within a generalised NED framework. We introduce a dimensionless scaling factor $\kappa$ as a phenomenological parameter modulating the nonlinear contributions in our model. This approach allows us to explore horizon stability and thermodynamic properties in a broader parameter space. In particular, we note that in the limiting case $\kappa=0$, the nonlinear corrections vanish, and the standard Reissner–Nordström (RN) solution is fully recovered. Overall, in our framework, while $\bar{\mu}_{\text{one-loop}}$ represents the physical QED correction, $\mu$ is the main parameter serving as the theoretical probe of the generalised nonlinear effects.\\

The metric function \eqref{nmetric} contains a sextic correction term of the form $\delta f(r) = -\frac{\bar{\mu} Q^4}{20 r^6}$, which is manifestly negative for $\bar{\mu} > 0$ and hence contributes as a short-range attractive potential near the black hole. This term becomes increasingly significant at small radial distances due to its $1/r^6$ dependence. Evaluating the metric function and its derivative for the maximal physically consistent values, one can confirm that this correction steepens the gradient of the metric function near the outer horizon. Since the Hawking temperature is proportional to the surface gravity, the enhanced slope directly translates into an increase in the black hole’s temperature. This result aligns with the expectation that stronger near-horizon gravitational effects (induced here by increasing $\bar{\mu}$) lead to enhanced thermal radiation. 
Therefore, within the physically viable parameter range, the $\bar{\mu}$-dependent correction acts as a short-range attractive modification that intensifies the gravitational field near the horizon and ,consequently, affects the thermodynamic profile of the black hole. This behaviour is explicitly illustrated in our revised temperature plots in Section \ref{sec3}.\vspace{0.7em}

Lastly, a few remarks can be made regarding the influence of the NED correction on the thermodynamic behaviour of the black hole.Throughout this section, it has been confirmed that the upper bounds on the charge and NED parameters are essential for maintaining physical consistency within the weak-field approximation. In particular, the impact of these constraints is clearly reflected in the Hawking temperature plots presented in the following section, where the thermodynamic behaviour aligns fully with theoretical expectations. We note, however, that if one were to consider field strengths approaching the Schwinger critical electric field—thus going beyond the weak-field regime—the effective upper bound on $\bar{\mu}$ could plausibly be higher. In such a scenario, the nonlinear corrections would become more pronounced, and the interplay between near-horizon gravitational strength and quantum evaporation could give rise to non-monotonic or unexpected thermodynamic behaviour. We regard this as an interesting direction for future research, particularly in strong-field extensions of the Euler-Heisenberg theory or in resummed formulations. For exploring this regime, the energy-momentum tensor given in \cite{ruffini2013einstein} for the case \(\varepsilon / E_c \gg 1\) may serve as a starting point for analysing the associated black hole structure.

\section{Thermodynamic Characteristics of Einstein-Euler-Heisenberg Black Hole} \label{sec3}

In this section, after the examination of the standard thermodynamic structure of the Einstein-Euler-Heisenberg black hole via Smarr formula\footnote{For a generalised Smarr formula and the first law of black hole thermodynamics (via covariant phase space formalism) for black holes with nonlinear electromagnetic fields, defined by electromagnetic Lagrangians dependent on both electromagnetic invariants (featuring Euler-Heisenberg and Born-Infeld Lagrangians), one may check \cite{gul,gul2}.}, the associated thermal fluctuations will be analysed from two different perspectives. Subsequently, the changes in the standard thermodynamic potentials and other variables will be calculated with two different corrections. In addition, the heat capacity which plays an important role in determining the stability for all cases will be investigated.
\subsection{The First Law of Thermodynamics and Smarr Formula}

To evaluate the thermodynamic parameters of the Einstein-Euler-Heisenberg black hole, we will be using the Smarr formula whose derivation can be carried out with the aid of Euler's homogeneous function theorem. With this purpose in mind, let us first express Euler's homogeneous function theorem \cite{wasserman2011thermal}

\begin{equation}
f(x_1,x_2,..,x_n)=l^{-1} \bigg[i\frac{\partial f}{\partial x_1}x_1+j\frac{\partial f}{\partial x_2}x_2+...+k\frac{\partial f}{\partial x_n}x_n\bigg],\label{X}
\end{equation}
where $\lambda$ is a constant, $(i,j,...,k)$ are the integer powers and $f$ represents a homogeneous function satisfying the homogeneity condition

\begin{equation}
f(\lambda^{i}x_1,\lambda^{j}x_2,...,\lambda^{k}x_n)=\lambda^{l}f(x_1,x_2,...,x_n).
\end{equation}

Now, let us find the mass of metric \eqref{nmetric}. To establish the connection between Euler's homogeneous function theorem and Smarr formula, one needs to set $r=r_H$ and equate the metric to function zero. As a result, the concerned mass function becomes
\begin{equation}
    M(r_{H},Q,\bar{\mu})= \frac{r_{H}}{2}+\frac{Q^{2}}{2r_{H}}- \frac{\bar{\mu} Q^{4}}{40 r_{H}^{5}}.\label{mass}
\end{equation}

Since for spherically symmetric black holes in four dimensions the Hawking-Bekenstein entropy in geometric units is 
\begin{equation}
S=\pi r^{2}_{H},\label{m6}
\end{equation}
one can alternatively express black hole mass \eqref{mass} as  
\begin{equation}
M(S,Q,\bar{\mu})=\frac{S^{1/2}}{2\pi^{1/2}}+\frac{Q^{2} \pi^{1/2}}{2S^{1/2}}-\frac{\bar{\mu} Q^4 \pi^{5/2}}{40S^{5/2}}.\label{m7}
\end{equation}

If the variables in \eqref{m7} are rewritten as

\begin{equation}
S \rightarrow \lambda^{i} S, \;\;\; \;  Q \rightarrow \lambda^{j} Q\;\;\; \; and  \;\;\; \;  \bar{\mu} \rightarrow \lambda^{k} \bar{\mu} 
\end{equation}
and the integer powers are set to satisfy

\begin{equation}
i=k=2l \;\;\;and \;\;\; \;  j=l,
\end{equation}

Euler's homogeneous function theorem \eqref{X} enables one to write

\begin{equation}
M(S,Q,\bar{\mu})=2S\left(\frac{\partial M}{\partial S}\right)+Q\left(\frac{\partial M}{\partial Q}\right)+2\bar{\mu} \left(\frac{\partial M}{\partial \bar{\mu}}\right).\label{m9}
\end{equation}
This equation, referred to as the Smarr formula, can alternatively be stated as

\begin{equation}
M=2\left(ST_{H}+\bar{\mu} \mathcal{A}\right)+Q\Phi_{e},
\end{equation}
where $\Phi_{e}$ and $\mathcal{A}$ represent the electric potential at the horizon and the conjugate quantity, respectively. By comparing Eqs.\eqref{X} and \eqref{m9}, the thermodynamic parameters of the Einstein-Euler-Heisenberg black hole can be determined. To start with, the Hawking temperature is evaluated as
\begin{equation}
T_{H}=\frac{\partial M}{\partial S}=\frac{1}{4\pi^{1/2}S^{1/2}}-\frac{Q^2\pi^{1/2}}{4S^{3/2}}+\frac{\bar{\mu} Q^4\pi^{5/2}}{16S^{7/2}},\label{mm10}
\end{equation}
whereas the electric potential at the horizon is found to be 
\begin{equation}
\Phi_{e}=\frac{\partial M}{\partial Q}=\frac{Q}{r_{H}}(1- \frac{\bar{\mu} Q^2}{10r_{H}^4}).\label{pot}
\end{equation}
The electric field  of the Einstein-Euler-Heisenberg black hole then becomes

\begin{equation}
    \mathcal{E}= - \nabla \Phi _{e} \hat{r}=\frac{Q}{r_{H}}(\frac{1}{r_{H}}-\frac{\mu Q^2}{2r_{H}^5})\; \hat{r}.
    \label{el}
\end{equation}

To investigate underlying scaling relations which might go unnoticed once thermodynamic variables are kept in dimensional form, one shall let $M \to M/M$, $r_H \to  r_h \equiv r_H/M$, $Q \to q \equiv  \; Q/M$ and $\bar{\mu} \to \mu \equiv  \; \bar{\mu} /M^2 $. Such an approach also ensures consistent comparability across black holes of varying masses. The same perspective will be maintained for all thermodynamic variables to be determined. Based on this perspective, Fig.\ref{wew} is plotted to show the behaviour of dimensionless potential\eqref{pot} and electric field\eqref{el} parallel to the potential for various $\mu$ values when the charge parameter is set to $q=1$. The plots suggest a correlation between the values of the horizon and the Euler-Heisenberg parameter at which the functions reach their peaks. Overall, the peaks are observed to reach higher values with decreasing $\mu$. For all cases, the effect of $\mu$ becomes irrelevant after  a certain horizon threshold.
\begin{figure}[H]
\centering
  \begin{tabular}{@{}cc@{}}
    \includegraphics{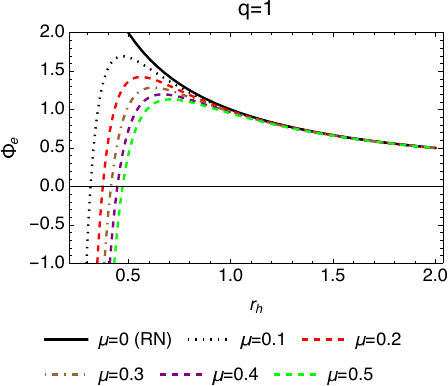}&    
    \includegraphics{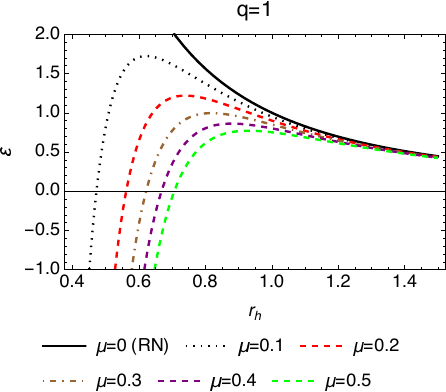} 
       \end{tabular}                 
 \caption{The plots above demonstrate the behaviour of the electric potential and the electric field against the event horizon, respectively, for $q=1$ and $\mu=0,0.1,0.2,0.3,0.4$ and $0.5$.}
\label{wew}
\end{figure}

The conjugate quantity is evaluated as follows.

\begin{equation}
\mathcal{A}=\frac{\partial M}{\partial \bar{\mu}}=-\frac{Q^4\pi^{5/2}}{40 S^{1/2}}.
\end{equation}

To express Hawking temperature in terms of horizon, one needs to  substitute Eq. \eqref{m6} into Eq. \eqref{mm10}. Doing so  results in

\begin{equation}
T_{H}=\frac{4r_{H}^6-4Q^2r_{H}^4+\bar{\mu} Q^4}{16\pi r_{H}^7}.\label{m10}
\end{equation}
This result can also be plotted for further analysis. In this context, Fig.\eqref{fig90} shows how the Hawking temperature behaves in relation to the horizon radius. From the graph it is evident that all plots converges in positive domain and the $\bar{\mu}=0$ case gives the behaviour of the Reissner Nordstrom black hole.

\begin{figure}[H]
\centering
  \begin{tabular}{@{}cccc@{}}
    \includegraphics{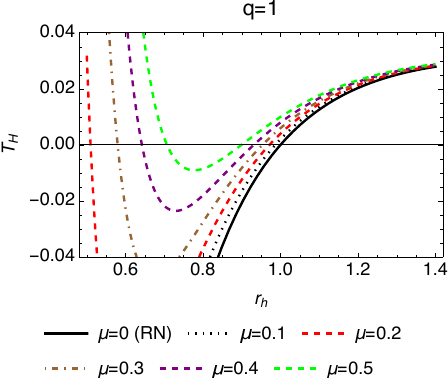} &
    \end{tabular}
 \caption{Hawking temperature versus horizon radius for $q=1$ and $\mu=0,0.1,0.2,0.3,0.4$ and $0.5$.}\label{fig90}
\end{figure}

Having determined the Hawking temperature, one can now derive other thermodynamic quantities that also play a vital role in describing the state of the concerned black hole.\\

The black hole's Helmholtz free energy can be written as 
\begin{equation}
F=-\int SdT_{H}=\frac{3Q^2}{4r_{H}}+\frac{r_{H}}{4}-\frac{7\bar{\mu} Q^4}{80r_{H}^5}.\label{s27}
\end{equation}
The internal energy, which determines the gravitational energy of the black hole and other thermodynamic interactions, is found as
\begin{equation}
E=\int T_{H} dS=\frac{r_{H}}{2}+\frac{Q^2}{2r_{H}}-\frac{\bar{\mu} Q^4}{40r_{H}^5}.
\end{equation}
\begin{figure}[H]
\centering
  \begin{tabular}{@{}cc@{}}
    \includegraphics{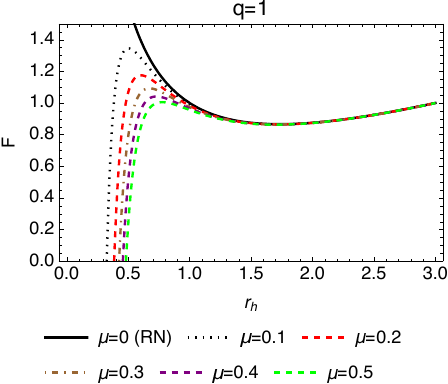}&    
    \includegraphics{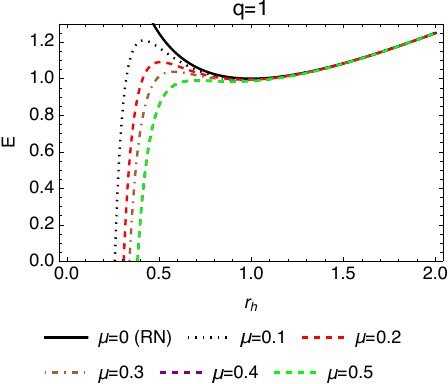} 
       \end{tabular}                 
 \caption{The Helmholtz free and internal energies are plotted for different NED parameter values for $q=1$. The effect of $\mu$ becomes negligible after a certain horizon value.} 
\label{fig42}
\end{figure}
Expressing pressure in terms of the horizon and the relevant NED parameters will help us in defining important thermodynamic potentials. Therefore, let us define the pressure of our thermodynamic system as 
\begin{equation}
    P=-\frac{dF}{dV}, \label{ss1}
\end{equation}
where $V=\frac{4}{3}\pi r^{3}_{H}$. Substituting Eq.\eqref{s27} into Eq.\eqref{ss1} , the pressure of the black hole becomes
\begin{equation}
    P=-\frac{dF}{dV}=-\frac{dF}{dr_{H}}\frac{dr_{H}}{dV}=\frac{1}{16\pi r^{2}_{H}}-\frac{3Q^{2}}{16\pi r^{4}_{H}}+\frac{7\bar{\mu} Q^{4}}{64r^{8}_{H}}. \label{ssss1}
\end{equation}
Setting the charge parameter as $q=1$, pressure \eqref{ssss1} is plotted as a function of horizon. The plots show that the black hole admits positive pressure for horizon values beyond  $1.8$. 

\begin{figure}[H]
\centering
  \begin{tabular}{@{}cccc@{}}
    \includegraphics{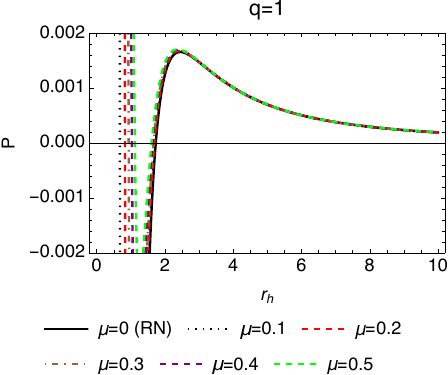} &

    \end{tabular}
 \caption{Pressure versus  black hole horizon for various $\mu$ values and fixed charge. }\label{figPR90}
\end{figure}

The enthalpy, which can be defined by taking the combination of other thermodynamic variables, is given by

\begin{equation}
H=E+PV. \label{s35}
\end{equation}

When  the related variables are plugged into Eq.\eqref{s35}, it reduces to

\begin{equation}
H=\frac{5r_{H}}{12}-\frac{Q^2}{4r_{H}}+\frac{29\bar{\mu} Q^4}{240r_{H}^{5} }  . \label{s36}
\end{equation}

Finally, the thermodynamic definition of Gibbs free energy is given by

\begin{equation}
G=F+PV. \label{s37}
\end{equation}

Upon substitution of variables, obtain Gibbs free energy is obtained as

\begin{equation}
G=\frac{Q^2}{2r_{H}}+\frac{r_{H}}{3}-\frac{7\bar{\mu} Q^4}{80r_{H}^5}+\frac{7\bar{\mu} Q^4}{48r_{H} }.
\end{equation}

\begin{figure}[H]
\centering
  \begin{tabular}{@{}cc@{}}
    \includegraphics{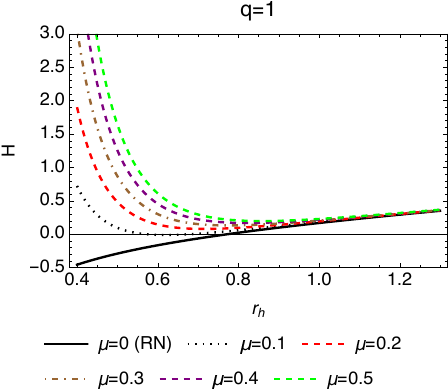}&    
    \includegraphics{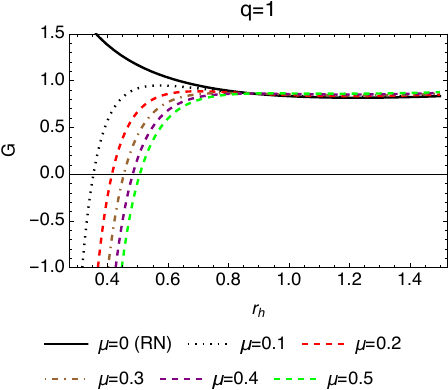} 
       \end{tabular}                 
 \caption{In the graphs above, the effect of the nonlinear parameter on the enthalpy and the Gibbs free energy of an Einstein-Euler-Heisenberg black hole is demonstrated for $q=1$. The peaks of the Gibbs free energy exist only for the cases when the NED effects are present.}
\label{figG90}
\end{figure}

Overall, the influence of the Euler-Heisenberg parameter on the thermodynamic quantities gets more distinguishable as the charge of the black hole increases.\\

Having determined the majority of thermodynamic parameters, one can now focus on deriving the heat capacity of the concerned black hole, which plays a significant role in examining the stability of the thermal system under consideration. In this regard, the heat capacity can be figured via 

\begin{equation}
C=T_{H}\left(\frac{\partial S}{\partial T_{H}}\right). \label{s25}
\end{equation}

Upon the substitution of Eq.\eqref{m10} and Eq.\eqref{m6} in Eq.\eqref{s25}, one obtains

\begin{equation}
C=-2 \pi r_{H}^2\left( \frac{4r_{H}^6-4Q^2r_{H}^4+\bar{\mu} Q^4}{4r_{H}^6-12Q^2r_{H}^4+7\bar{\mu} Q^4}  \right). \label{s26}
\end{equation}

\begin{figure}[H]
\centering
  \begin{tabular}{@{}cccc@{}}
    \includegraphics{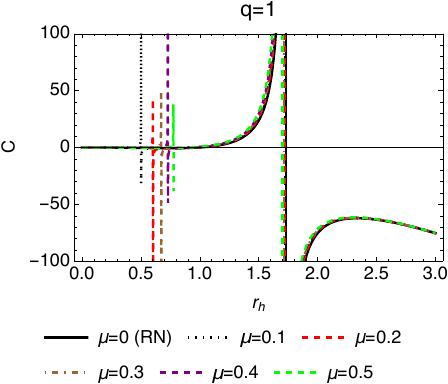} &

    \end{tabular}
 \caption{The heat capacity versus event horizon graphs according to Eq.\eqref{s26} for different NED parameter values.}\label{fig1}
\end{figure}

The plots presented in Fig.\ref{fig1} show the behaviour of the heat capacity $C$ as a function of $r_h$, under different choices of $\mu$. To assess the physical properties of the concerned black hole, one can impose the upper bounds \(\mu = 50/81\) and \(q = \sqrt{25/24}\) necessary to ensure consistency with the weak-field approximation. Under these conditions, the behaviour of the heat capacity indicates the existence of a second-order phase transition: the denominator of the expression vanishes at a critical horizon radius \(r_c \approx 1.7\), where the heat capacity diverges (as can also be seen from the figure). This marks a transition point between two thermodynamic phases. For \(r_h < r_c\), the heat capacity is positive, indicating a thermodynamically stable branch of small black holes. In contrast, for \(r_h > r_c\), the heat capacity becomes negative and the black hole enters an unstable phase. This result demonstrates that the inclusion of the Euler-Heisenberg correction allows for the existence of a locally stable phase at small scales, bounded by a critical radius beyond which thermodynamic instability resumes. The critical behaviour observed here is characteristic of phase transitions in black hole thermodynamics and confirms that the NED modification plays a stabilising role within a restricted domain of the parameter space. A graphical representation of this behaviour is provided in Fig.\ref{fig1}, where the divergence and stability regimes can be observed.

In the case where the black hole entropy takes the classical Bekenstein–Hawking form, the stability landscape exhibits a relatively simple and symmetric structure. As illustrated in Fig.\ref{figgg1}, the stability region appears as a broad horizontal band, primarily determined by the electric charge parameter \(q\). This region extends approximately within the range \(0.8 \lesssim q \lesssim 1.2\), and is largely insensitive to variations in the NED parameter \(\mu\), which confirms that nonlinear corrections play a minimal role in this baseline scenario.

Thermodynamic stability, as indicated by the positivity of the heat capacity, is thus governed mainly by the balance between gravitational attraction and electromagnetic repulsion. For charges too low (\(q \lesssim 0.8\)), repulsion is insufficient to counteract the gravitational collapse, leading to instability. Conversely, at higher charges (\(q \gtrsim 1.2\)), the repulsive force becomes dominant, destabilising the configuration.

Importantly, within the physically admissible region determined by our theoretical bounds—namely \(\mu \leq \tfrac{50}{81} \approx 0.617\) and \(q \leq \sqrt{25/24} \approx 1.021\)—a portion of the stability band does persist. This confirms that thermodynamically stable black hole solutions are not only mathematically admissible, but also physically viable under the assumptions of weak-field Euler--Heisenberg theory. The intersection of this narrow stable zone with the constrained parameter space plays a crucial role in validating the consistency of the model and its semi-classical interpretation.

The resulting stability map is marked by a single continuous phase transition boundary, corresponding to a second-order transition where the heat capacity diverges. This uncorrected case does not exhibit additional internal structure or parameter interdependence. The thermodynamic behaviour here is therefore consistent with classical expectations, offering a clean benchmark against which the quantum-corrected scenarios may be contrasted.

\begin{figure}[H]
\centering
  \begin{tabular}{@{}cccc@{}}
    \includegraphics[width=7cm]{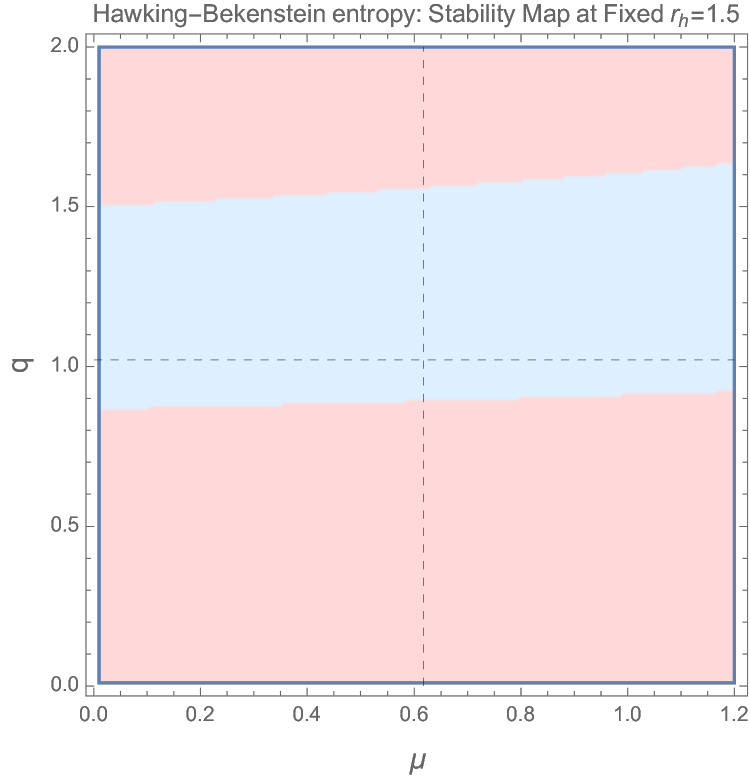} &

    \end{tabular}
 \caption{Stability map for the black hole model with classical Bekenstein--Hawking entropy, plotted in the \((\mu, q)\) parameter space at fixed horizon radius \(r_h = 1.5\). The blue region denotes thermodynamically stable configurations where the heat capacity is positive, while the red region corresponds to instability. The black dashed lines indicate the upper bounds \(\mu = \tfrac{50}{81}\) and \(q = \sqrt{25/24}\), which define the physically admissible domain under the weak-field approximation. A single, nearly horizontal phase transition boundary separates the stable and unstable zones, with stability predominantly determined by the electric charge parameter \(q\) and only weakly influenced by \(\mu\)}\label{figgg1}
\end{figure}

\subsection{ Corrections of Thermodynamic Parameters Based on Entropy}

The thermodynamic variables under examination may undergo intriguing modifications  if statistical mechanical corrections are added to the thermodynamic studies. In this framework, the fluctuations of the spacetime under study can be examined together with their impact on the macro scale properties of the system via statistical physics. With this purpose, let us use stability analysis to inspect the impact of logarithmic and exponential variations on the thermodynamics of Einstein-Euler-Heisenberg.

\subsubsection{Logarithmic Corrections to Black Hole Entropy}

Using the distribution function defined in \cite{gibbons1977action}, one may start investigating the logarithmic correction to the entropy based on definition of the partition function from the statistical mechanical perspective that goes as follows.

\begin{equation}
Z=\int _0^\infty \rho(\sigma) e^{-\beta \sigma}d\sigma \label{c1},
\end{equation} 

in which $\beta=1/T_H$ and $\sigma$ represent the thermal average energy. To determine the density of states, which is the inverse Laplace transform of  Eq.\eqref{c1} for a certain energy, one can use

\begin{equation}
\rho=\frac{1}{2\pi i}\int _{c-i\infty}^{c+i\infty}  e^{\mathcal{S}(\beta)}d\beta \label{c2},
\end{equation} 

where $S(\beta)=\beta\sigma+\ln Z$. To perform the above integral perturbatively using the saddle-point method, let us expand, at least in principle, the series expansion of entropy expression in the exponential term about the saddle-point $\beta_0$ can be written as

\begin{equation}
\mathcal{S}(\beta)\approx S_0+\frac{1}{2}\left(\beta-\beta_0 \right)^2\left(\frac{\partial^2\mathcal{S}}{\partial\beta^2} \right)_{\beta_0}+..., \label{c3}
\end{equation} 

in which $S_0$ is the zeroth order entropy. When  Eq.\eqref{c3} is substituted into Eq.\eqref{c2}, the density of states becomes

\begin{equation}
\rho=\frac{e^{S_0}}{2\pi i}\int _{c-i\infty}^{c+i\infty}  e^{\frac{1}{2}\left(\beta-\beta_0 \right)^2\left(\frac{\partial^2\mathcal{S}}{\partial\beta^2} \right)_{\beta_0}}d\beta \label{c4}.
\end{equation} 

Consequently, the result of integral \eqref{c4} comes out as \cite{das2002general}

\begin{equation}
\rho=\frac{e^{S_0}}{\sqrt{2\pi \left(\frac{\partial^2\mathcal{S}}{\partial\beta^2} \right)_{\beta_0} }} \label{c5}.
\end{equation} 

Eventually, the microcanonical entropy at equilibrium can be written as 

\begin{equation}
S_{C}=ln\rho=S_{0}-\frac{k}{2} ln S_{0}T_{H}^{2},\label{s39}
\end{equation} 

where  $k$ is the thermal fluctuation parameter \cite{faizal2015correction}. In this work, the  thermal fluctuation parameter will be taken as $k=1$ considering that the contribution from these fluctuations is at its highest when $k=1$. If  Hawking temperature \eqref{m10} and the zeroth order entropy $\pi r_h^2$ are plugged in Eq.\eqref{s39}, the microcanonical entropy of the system becomes

\begin{equation}
S_{C}=\pi r_H^2-\frac{1}{2}ln\left( \frac{(4r_{H}^6-4Q^2r_{H}^4+\bar{\mu} Q^4)^2}{256\pi r_{H}^{12}} \right) . \label{s40}
\end{equation}

The corrections on the internal energy of the system can be calculated using the general formula defined as

\begin{equation}
E_{C}=\int T_{H} dS_C. \label{s41}
\end{equation}

substituting corrected entropy \eqref{s40} and Hawking temperature \eqref{m10} in Eq.\eqref{s41}, one finds

\begin{equation}
\begin{aligned}
E_{C}=\frac{420 \pi Q^{2} r_{H}^{6}+420 \pi r_{H}^{8}-21 \bar{\mu} \pi Q^{4} r_{H}^{2}+140  Q^{2} r_{H}^{4}-45 \bar{\mu} q^{4}}{840 \pi r_{H}^{7}}.
\end{aligned} \label{s42}
\end{equation}

The Helmholtz free energy can be modified using the logarithmically corrected entropy as follows.

\begin{equation}
F_{C}=-\int S_{C} dT_{H}. \label{s43}
\end{equation}

Plugging Eq.\eqref{s40} and Eq.\eqref{m10} into Eq.\eqref{s43}, one finds

\begin{equation}
\begin{aligned}
    F_{C} = -\frac{1}{3360\pi r_{H}^7} \Big[ &
    3 \bar{\mu} Q^4 \left(98 \pi r_{H}^2 + 60 + 35 \ln(256 \pi)\right)  \\
    & - 105 \left(\bar{\mu} Q^4 - 4 Q^2 r_{H}^4 + 4 r_{H}^6\right) \ln \left(\frac{\left(\bar{\mu} Q^4 - 4 Q^2 r_{H}^4 + 4 r_{H}^6\right)^2}{r_{H}^{12}}\right) \\
    & - 140 Q^2 r_{H}^4 \left(18 \pi r_{H}^2 + 4 + 3 \ln(256 \pi)\right) \\
    &  + 420 r_{H}^6 \left(-2 \pi r_{H}^2 + \ln(256) + \ln(\pi)\right) 
    \Big].
\end{aligned}
\end{equation}

\begin{figure}[H]
\centering
  \begin{tabular}{@{}cc@{}}
    \includegraphics{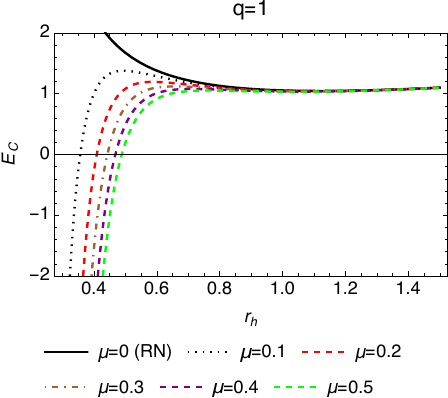}&    
    \includegraphics{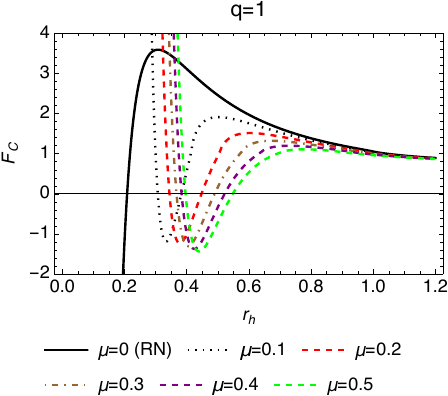} 
       \end{tabular}                 
 \caption{Logarithmically corrected internal and Helmholtz free energies versus black hole horizon for various  $\mu$ values.  Once the Helmholtz free energy graph is compared with the uncorrected ones, (see Fig.\ref{fig42}) one notices that the trends are the exact opposite. }  
\label{fig3}
\end{figure}

Fig.\eqref{fig3} illustrates the behaviour of the logarithmically corrected internal and Helmholtz free energies  as a function of $r_h$, under different  nonlinear parameters. In the first plot, each curve represents the variation of $E_c$ with $r_h$ whose overall evaluation shows an increase in the charge parameter leads to a rise in the peak value of the logarithmically corrected internal energy, while an increase in the nonlinear parameter causes a decrease in the concerned peaks. As the value of $r_h$ increases, the effect of  the NED parameter becomes irrelevant and all functions coincide regardless of the charge parameter. In the second plot of Fig.\eqref{fig3} , as in the logarithmically corrected internal energy, the influence of different  $\mu$ values become apparent at small $r_h$ values. Increasing  the charge causes the values of the minimum values of corrected free energy to decrease. Also, increasing the nonlinear parameter values results in a shift of the minimum points towards higher negative values. When the $r_h$ value increases, the contribution of the nonlinear parameter disappears.

After finding the logarithmically corrected Helmholtz free energy, the logarithmically corrected pressure can be calculated via

\begin{equation}
P_{C}=-\frac{dF_{C}}{dV},\label{pressure}
\end{equation}

which results in

\begin{equation}
\begin{aligned}
    P_{C} = \frac{1}{128 r_{H}^{10} \pi^{2}} \Big[ &
    (12 Q^{2} r_{H}^{4} - 4 r_{H}^{6} - 7 \bar{\mu} Q^{4}) \ln\left(\frac{(-4 Q^{2} r_{H}^{4} + 4 r_{H}^{6} + \bar{\mu} Q^{4})^{2}}{r_{H}^{12}}\right) \\
    & + (-12 Q^{2} r_{H}^{4} + 4 r_{H}^{6} + 28 \bar{\mu}) \ln(\pi) \\
    & + (-96 Q^{2} r_{H}^{4} + 32 r_{H}^{6} + 224 \bar{\mu}) \ln(2) \\
    & + 8 r_{H}^{8} \pi - 24 Q^{2} r_{H}^{6} \pi + 56 \bar{\mu} \pi r_{H}^{2} + (-12 Q^{4} + 48) \bar{\mu} \Big].
\end{aligned}
\end{equation}

\begin{figure}[H]
\centering
  \begin{tabular}{@{}cccc@{}}
    \includegraphics{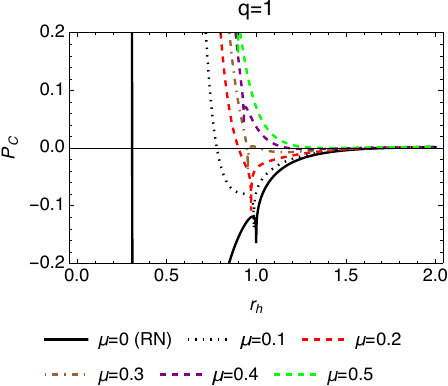} &

    \end{tabular}
 \caption{Logarithmically corrected pressure versus the black hole horizon. The negative pressure values might give indications of a repulsive effect arising due to vacuum polarisation. Once all the pressure values coincide at zero, the black hole reaches an equilibrium state.}\label{fig0}
\end{figure}
As can be seen in Fig.\eqref{fig0}, increasing the  NED parameter values result in a black hole with negative pressure. This might suggest a resistance towards a possible collapse of the black hole.\\ 

Using the logarithmically corrected pressure and internal energy, Eq.\eqref{s35}  becomes

\begin{equation}
\begin{aligned}
    H_{C} = \frac{1}{3360 \pi r_{H}^7} \Big[ & 
    (420 Q^{2} r_{H}^{4} - 140 r_{H}^{6} - 245 \bar{\mu} Q^{4}) \ln\left(\frac{(-4 Q^{2} r_{H}^{4} + 4 r_{H}^{6} + \bar{\mu} Q^{4})^{2}}{r_{H}^{12}}\right) \\
    & + (-420 Q^{2} r_{H}^{4} + 140 r_{H}^{6} + 980 \bar{\mu}) \ln(\pi) \\
    & + (-3360 Q^{2} r_{H}^{4} + 1120 r_{H}^{6} + 7840 \bar{\mu}) \ln(2) \\
    & + 1960 \pi r_{H}^{8} + 840 \pi Q^{2} r_{H}^{6} \\
    & + 560  Q^{2} r_{H}^{4} - 84 \left(Q^{4} - \frac{70}{3}\right) \pi \bar{\mu} r_{H}^{2} \\
    & - 180 \bar{\mu} \left(-\frac{28}{3} +  \frac{10}{3} Q^{4}\right) \Big],
\end{aligned}
\end{equation}

which represents the logarithmically corrected  enthalpy. Also, the  logarithmically corrected Gibs free energy  is determined as

\begin{equation}
\begin{aligned}
    G_{C} = \frac{1}{840 \pi r_{H}^{7}} \Big[ &
    420 \pi Q^{2} r_{H}^{6} + 280 \pi r_{H}^{8} + 70 \ln\left(\frac{(-4 Q^{2} r_{H}^{4} + 4 r_{H}^{6} + \bar{\mu} Q^{4})^{2}}{r_{H}^{12}}\right) r_{H}^{6} \\
    & - 70 \ln(\pi) r_{H}^{6} - 560 \ln(2) r_{H}^{6} - 35 \ln\left(\frac{(-4 Q^{2} r_{H}^{4} + 4 r_{H}^{6} + \bar{\mu} Q^{4})^{2}}{r_{H}^{12}}\right) \bar{\mu} Q^{4} \\
    & + 140 Q^{2} r_{H}^{4} - 105 \bar{\mu} Q^{4} + 196 \bar{\mu} \pi r_{H}^{2} + 140 \ln(\pi) \bar{\mu} + 1120 \ln(2) \bar{\mu} + 240 \bar{\mu} \Big],
\end{aligned}
\end{equation}

following Eq.\eqref{s37}.

\begin{figure}[H]
\centering
  \begin{tabular}{@{}cc@{}}
    \includegraphics{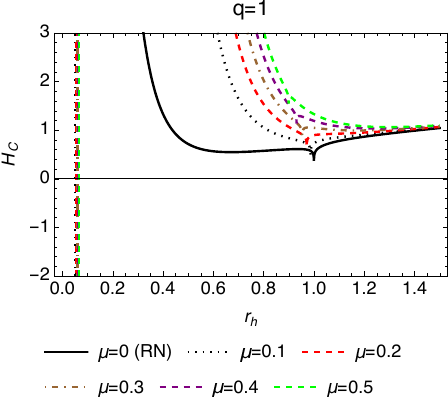}&    
    \includegraphics{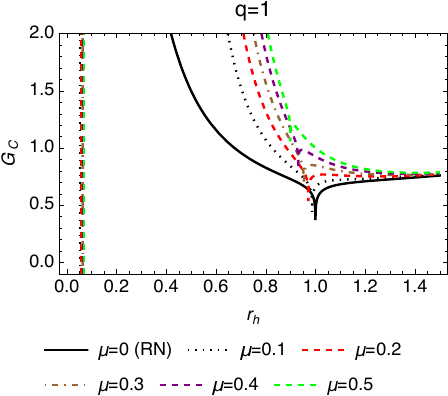} 
       \end{tabular}                 
 \caption{Logarithmically corrected enthalpy and  Gibs free energy  versus black hole horizon. Logarithmically  corrected Gibbs free energy versus black hole horizon for various $q$ and $\mu$ values. }  
\label{fig6n}
\end{figure}

The behaviour of the logarithmically corrected enthalpy and  Gibs free energy  in relation to the event horizon radius is shown in Fig.\eqref{fig6n}. The first plot shows that the gradient of the corrected enthalpy declines behaviour with increasing radius, eventually converging to a single function. For all charge parameter values, the corrected enthalpy roughly behaves as reflections of its uncorrected versions through the x-axis. The second in Fig.\ref{fig6n} is displayed to investigate the behaviour of the logarithmically corrected  Gibbs free energy and the impact of thermal variations. When these graphs are compared with the ones given in Fig.\ref{figG90}, one notices a drastic change in the behaviour of the peak points. To be more specific, the peaks observed in  Fig.\ref{fig6n} are sharp, while those in Fig.\ref{figG90} experience a smooth variations. \\

To define the logarithmically  corrected  heat capacity, one must use 

\begin{equation}
C_{C}=\frac{d E_{C}}{dT_{H}}\label{stability},
\end{equation}

which becomes

\begin{equation}
    C_{C}= \frac{2 \bar{\mu} \pi Q^4 r_H^2 + 6 \bar{\mu} Q^4 - 8 \pi Q^2 r_H^6 + 8 \pi r_H^8 - 8 Q^2 r_H^4}{-7 \bar{\mu} Q^4 + 12 Q^2 r_H^4 - 4 r_H^6}.
\end{equation}

\begin{figure}[H]
\centering
  \begin{tabular}{@{}cccc@{}}
    \includegraphics{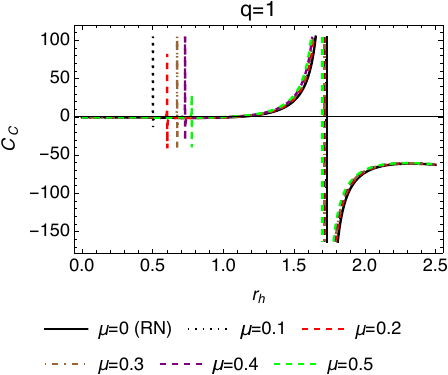} &

    \end{tabular}
 \caption{Logarithmically corrected heat capacity versus the black hole horizon for various $\mu$ values.}\label{Fig1}
\end{figure}

In the presence of logarithmic corrections to the black hole entropy, the thermodynamic stability structure becomes more constrained in comparison to the classical case. As shown in Fig.\ref{figgg2}, the stable region remains broadly similar in shape—retaining a horizontal band around \(q \sim 1\)—but its width is significantly reduced. This results in a noticeably thinner zone of thermodynamic stability. The corrected heat capacity remains positive only within a more limited corridor of the \((\mu, q)\) parameter space.

Despite this narrowing, the overall qualitative features of the stability landscape remain consistent with the Bekenstein-Hawking baseline. The stability band continues to exhibit only weak dependence on the NED parameter, suggesting the logarithmic correction primarily influences the threshold behaviour along the \(q\)-direction. This indicates that the leading-order quantum correction alters the effective energy balance between gravitational and electromagnetic contributions without introducing fundamentally new phase structure.

Of particular relevance is the fact that a segment of the stability region still lies within the physically admissible window defined by \(\mu \leq \tfrac{50}{81}\) and \(q \leq \sqrt{25/24}\). This ensures that, even with the inclusion of logarithmic quantum corrections, the model admits thermodynamically stable black hole configurations that are consistent with the weak-field approximation and semi-classical expectations. However, the reduced size of this viable region highlights the sensitivity of stability to subleading quantum corrections.

The phase transition boundary, marked by the condition \(C = 0\), continues to be defined by a single critical curve in the \((\mu, q)\) plane. While less pronounced than in the exponential case, this boundary plays a central role in delimiting the permissible thermodynamic phase space and illustrates how even modest corrections to the entropy can influence the macroscopic behaviour of the system.

\begin{figure}[H]
\centering
  \begin{tabular}{@{}cccc@{}}
    \includegraphics[width=7cm]{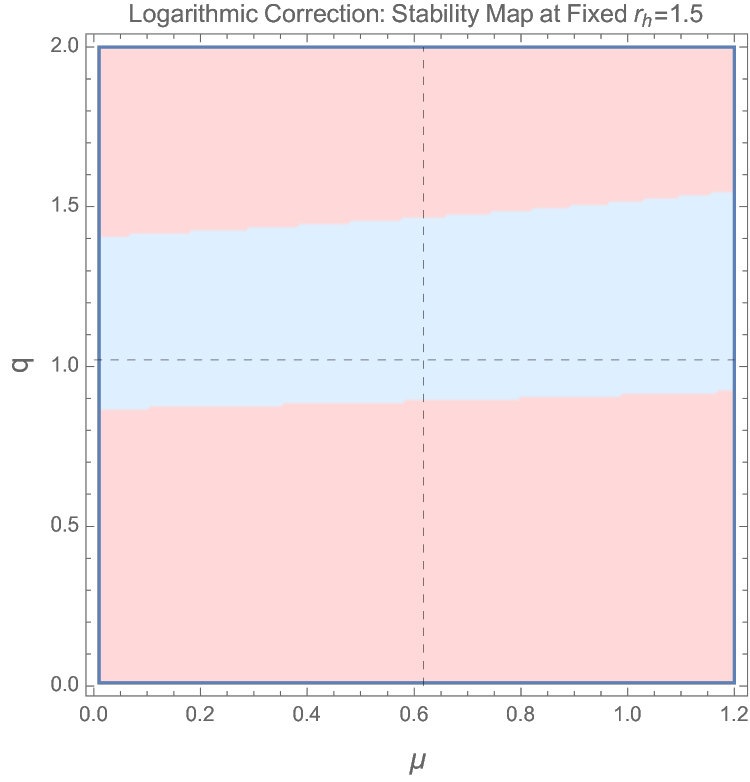} &

    \end{tabular}
 \caption{Stability map for the black hole model incorporating logarithmic corrections to the entropy. As in the classical case, the phase diagram is presented in the \((\mu, q)\) space for fixed \(r_h = 1.5\). The blue region identifies stable configurations with positive heat capacity, and the red region indicates instability. The stability band is notably thinner than in the uncorrected case, though it retains a similar horizontal structure with minimal sensitivity to \(\mu\). The admissible region bounded by \(\mu = \tfrac{50}{81}\) and \(q = \sqrt{25/24}\) intersects part of the stable zone, confirming the physical viability of solutions in the presence of logarithmic entropy corrections.}\label{figgg2}
\end{figure}

\subsubsection{Exponential Corrections to Black Hole Entropy}

The exponential correction to the black hole entropy in  Euler-Heisenberg NED when can be discuss  microstate counting is done for quantum states that are only present on the horizon. In this case, the entropy of the black hole receives an exponential correction \cite{chatterjee2020exponential}. The exponential statistical correction can be applied by treating the  black hole as a system  consisting of a set of $N$ micro-particles. In this context, the number of different permutations of the elements in the statistical ensemble is given as \cite{chatterjee2020exponential}

\begin{equation}
    \Omega=\frac{\left(\sum s_i\right)!}{\prod s_i!},
\end{equation}
in which $s_i$ the is the repetition number of each particle. Also note that when $n_i$ and $\mathcal{\epsilon}_i$ are the number of each particle and the energy of the $i$ th microstate, respectively, the total number of micro-particles and the total energy of the statistical system can be written as

\begin{equation}
\begin{aligned}
&N=\sum s_in_i,\\
&E=\sum s_in_i\mathcal{\epsilon}_i.
\end{aligned}
\end{equation}
The variation of $\ln\Omega$  under the constraints $\delta\sum s_in_i=0$ and $\sum \mathcal{\epsilon}_i\delta\left(s_in_i \right)=0$ gives the repetition number of the particles as

\begin{equation}
s_i=\left(\sum s_i \right)e^{-\lambda n_i},
\end{equation}
where $\lambda$ is the variation parameter. Therefore, the constraint $\sum e^{-\lambda n_i}=1$, the large  $N$  approximation and the entropy definition  $S=\lambda N$ all together enable one to define the exponentially corrected entropy as \cite{chatterjee2020exponential}

\begin{equation}
    S^{EC}= S_0 + e^{-S_0}\label{EC1}.
\end{equation}
Starting from the following definition the exponentially  internal energy and substituting  Hawking temperature \eqref{m10} in Eq. \eqref{s41}, one gets

\begin{equation}
     E^{EC}=\frac{1}{8} \left(\text{erfi}\left(\sqrt{\pi } r_{H}\right) \left(\frac{4}{15} \pi ^3 \bar{\mu}  Q^4-4 \pi  Q^2+2\right)-\frac{\bar{\mu}  Q^4 \left(e^{\pi  r_{H}^2} \left(4 \pi ^2 r_{H}^4+2 \pi  r_{H}^2+3\right)+3\right)}{15 r_{H}^5}+\frac{4 Q^2 \left(e^{\pi  r_H^2}+1\right)}{r_{H}}+4 r_{H}\right),\label{zz4}
\end{equation}
in which $\text{``erfi"}$ represents the error function. The exponentially corrected Helmholtz free energy  becomes

\begin{equation}
\begin{aligned}
    F^{EC} = \frac{1}{240 \pi r_{H}^7} \Big[ 
    & 4 \pi r_{H}^7 \text{erfi}\left(\sqrt{\pi} r_{H}\right) \left(2 \pi^3 \bar{\mu} Q^4 - 30 \pi Q^2 + 15\right) \\
    & + 3 \pi r_{H}^2 \left(-7 \bar{\mu} Q^4 + 60 Q^2 r_{H}^4 + 20 r_{H}^6\right) \\
    & + e^{\pi r_H^2} \left(60 r_{H}^4 \left(\left(2 \pi Q^2 - 1\right) r_{H}^2 + Q^2\right) - \bar{\mu} Q^4 \left(8 \pi^3 r_{H}^6 + 4 \pi^2 r_{H}^4 + 6 \pi r_{H}^2 + 15\right)\right) \Big],\label{zz3}
\end{aligned}
\end{equation}

following the same procedure as the  exponentially corrected internal enargy.

\begin{figure}[H]
\centering
  \begin{tabular}{@{}cc@{}}
    \includegraphics{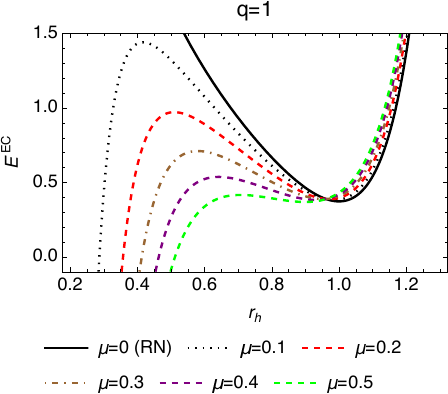}&    
    \includegraphics{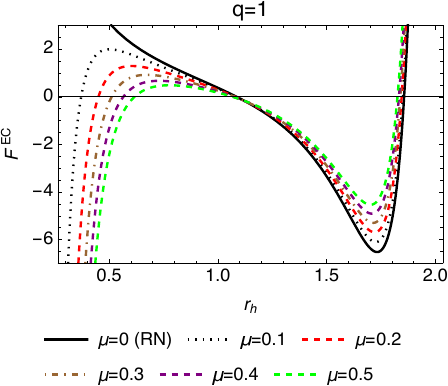} 
       \end{tabular}                 
 \caption{Exponentially corrected  internal and Helmholtz free energies versus  black hole horizon for various $\mu$ values.}  
\label{fig52}
\end{figure}

Fig.\eqref{fig52} shows the evolution of the exponentially corrected  internal and Helmholtz free energies against $r_h$ for different charge and nonlinear parameter values. The main difference between the first plot and the ones drawn for the uncorrected and the logarithmically corrected internal energies can be seen for $r_h>1$ when the charge parameters is $q=1$. Regardless of  the charge parameter values, all functions start divergent after the mentioned threshold horizon values towards a preferred direction depending on the NED value. On the other hand, when the exponentially corrected Helmholtz free energy is compared with the uncorrected (Fig.\ref{fig42})  and logarithmically corrected (Fig.\ref{fig3}) ones, one notices that only for the exponentially corrected case all functions diverges.

If one inserts Eq.\eqref{zz3} into Eq.\eqref{pressure}, the  exponentially corrected pressure can be explicitly written as  

\begin{equation}
\begin{aligned}
    P^{EC} = \frac{1}{960 \pi^2 r_{H}^{10}} \Big[ 
    & 8 \pi r_{H}^7 \text{erfi}\left(\sqrt{\pi} r_{H}\right) \left(2 \pi^3 \bar{\mu} Q^4 - 30 \pi Q^2 + 15\right) \\
    & - 147 \pi \bar{\mu} Q^4 r_{H}^2+ 60 \pi r_{H}^6 \left(9 Q^2 + r_{H}^2\right) \\
    & + e^{\pi r_{H}^2} \left(60 r_{H}^4 \left(\left(4 \pi Q^2 - 3\right) r_{H}^2 + 5 Q^2\right) - \bar{\mu} Q^4 \left(4 \pi \left(4 \pi^2 r_{H}^4 + 2 \pi r_{H}^2 + 3\right) r_{H}^2 + 135\right)\right) \Big].\label{zz5}
\end{aligned}
\end{equation}

\begin{figure}[H]
\centering
  \begin{tabular}{@{}cccc@{}}
    \includegraphics{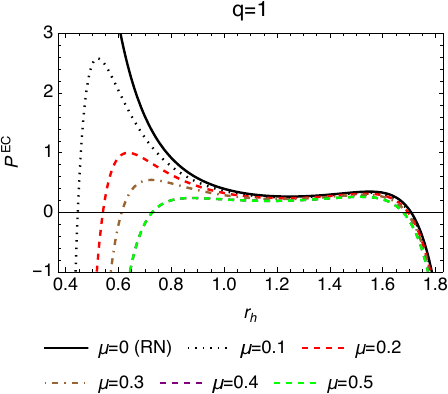} &
    \end{tabular}
 \caption{Exponentially corrected pressure versus black hole horizon for various $\mu$ values.}\label{Fig53}
\end{figure}

Fig.\eqref{Fig53} illustrates how the exponentially corrected pressure shifts from  positive to negative values as $r_h$ increases. The black hole and switches into attractive or repulsive behaviour depending on the NED parameter.\\

When we put Eq.\eqref{zz4} and Eq.\eqref{zz5} into Eq.\eqref{s35}, the exponentially corrected enthalpy can be written as 

\begin{equation}
\begin{aligned}
    H^{EC} = \frac{1}{144 \pi r_{H}^7} \Big[ 
    & 4 \pi r_{H}^7 \text{erfi}\left(\sqrt{\pi} r_{H}\right) \left(2 \pi^3 \bar{\mu} Q^4 - 30 \pi Q^2 + 15\right) \\
    & + 3 \pi r_H^2 \left(-11 \bar{\mu} Q^4 + 60 Q^2 r_{H}^4 + 28 r_{H}^6\right) \\
    & - e^{\pi r_{H}^2} \left(\bar{\mu} Q^4 \left(8 \pi^3 r_{H}^6 + 4 \pi^2 r_{H}^4 + 6 \pi r_{H}^2 + 27\right) - 60 Q^2 \left(2 \pi r_{H}^6 + r_{H}^4\right) + 36 r_{H}^6\right) \Big].
\end{aligned}
\end{equation}

One of the state functions  essential for determining the global stability of the black hole is the Gibbs free energy. For an Einstein-Euler-Heisenberg black hole, the relation for Gibbs free energy can be found via

\begin{equation}
\begin{aligned}
    G^{EC} = \frac{1}{72 \pi r_{H}^7} \Big[ 
    & 2 \pi r_{H}^7 \text{erfi}\left(\sqrt{\pi} r_{H}\right) \left(2 \pi^3 \bar{\mu} Q^4 - 30 \pi Q^2 + 15\right) \\
    & + 3 \pi r_{H}^2 \left(-7 \bar{\mu} Q^4 + 36 Q^2 r_{H}^4 + 8 r_{H}^6\right) \\
    & - e^{\pi r_{H}^2} \left(\bar{\mu} Q^4 \left(\pi \left(4 \pi^2 r_{H}^4 + 2 \pi r_{H}^2 + 3\right) r_{H}^2 + 18\right) - 12 Q^2 \left(5 \pi r_{H}^2 + 4\right) r_{H}^4 + 36 r_{H}^6\right) \Big].
\end{aligned}
\end{equation}

\begin{figure}[H]
\centering
  \begin{tabular}{@{}cc@{}}
    \includegraphics{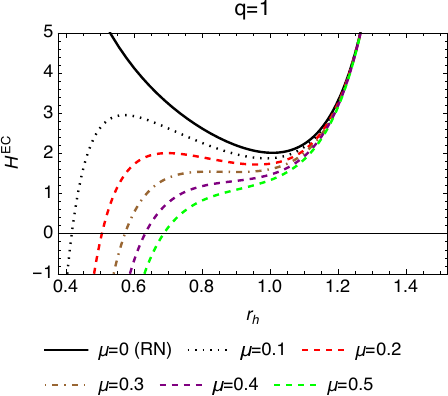}&    
    \includegraphics{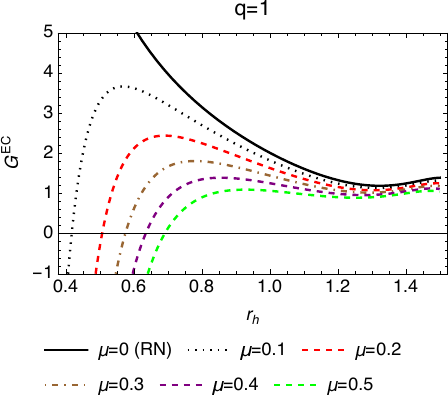} 
       \end{tabular}                 
 \caption{Exponentially corrected enthalpy and  Gibbs free energy versus the black hole horizon for various $\mu$ values.}  
\label{fig555}
\end{figure}

In the first plot of Fig.\ref{fig555} , all enthalpy positive beyond $r_h=0.7$ as well as in the second plot of Fig.\eqref{fig555}.

Finally, heat capacity, one of the most important thermodynamic quantities, can be examined. To see the effect of the exponential modification of heat capacity, one can put $E^{EC}$ into  Eq.\eqref{stability}, which leads to
\begin{equation}
\begin{aligned}
C^{EC} = \frac{16 \pi r_H^2 e^{\pi r_H^2}}{7 \bar{\mu} Q^4 - 12 Q^2 r_H^4 + 4 r_H^6} & \left[ \bar{\mu} \pi Q^4 r_H^2 \left(0.0667 \pi r_H^4 + 0.0333 \pi r_H^2 + 0.1333 r_H^2 + 0.0833 \right) \right. \\
& - 0.0417 \bar{\mu} Q^4 \left(4 \pi r_H^4 + 2 \pi r_H^2 + 3 \right) 
 + 0.5 Q^2 r_H^4 \left(e^{\pi r_H^2} + 1 \right) \\
& - r_H^6 \left(1.0 \pi Q^2 + 0.5 \right) 
 - \left. r_H^6 \left(0.0667 \bar{\mu} \pi^3 Q^4 - Q^2 + 0.5 \right) e^{\pi r_H^2} - 0.125 \right].
\end{aligned}
\end{equation}

\begin{figure}[H]
\centering
  \begin{tabular}{@{}cccc@{}}
    \includegraphics{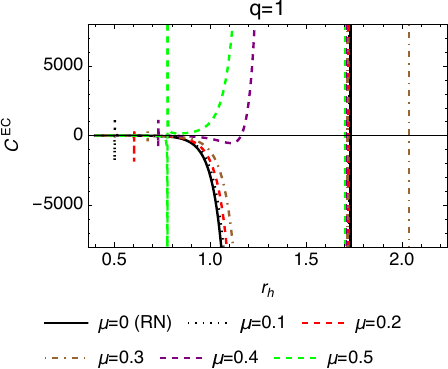} &
    \end{tabular}
 \caption{Exponentially corrected stability  versus the black hole horizon for various $q$ and $\mu$ values.}
 \label{Fig2}
\end{figure}

The plots presented in Fig.\ref{Fig2} represent the behaviour of  $C^{EC}$ as a function of $r_h$, under different $\mu$ parameters.

When exponential corrections are incorporated to the black hole entropy, the thermodynamic stability structure undergoes a significant qualitative transformation. As shown in Fig.\ref{figgg3}, the resulting stability region no longer forms a simple horizontal band as in the uncorrected and logarithmic cases. Instead, it displays a distinctly curved and structured form, with stability now depending strongly on both the NED parameter \(\mu\) and the electric charge \(q\).

This richer structure arises from the enhanced influence of quantum vacuum effects encoded in the exponential correction. In particular, the stable region tends to occupy zones with moderate values of \(\mu\) and slightly elevated values of \(q\), with the boundaries of this region bending inward or outward depending on the specific interplay between these parameters. The heat capacity becomes highly sensitive to small changes in \(\mu\) and \(q\), indicating that the exponential term induces a more complex energy balance between gravitational and electromagnetic contributions.

Notably, the physically admissible window defined by \(\mu \leq \tfrac{50}{81}\) and \(q \leq \sqrt{25/24}\) intersects with only a narrow segment of the stable region. This implies that, while stable solutions are still present under physically consistent assumptions, they are more finely tuned and restricted compared to the previous cases. In some portions of the admissible domain, even small deviations in \(\mu\) or \(q\) can drive the system into thermodynamic instability.

The phase transition boundary, defined by the condition \(C = 0\), takes on a more elaborate form. Unlike the single, nearly linear boundary observed in the uncorrected case, the exponential model yields a curved and branching transition contour, delineating multiple zones of stability and instability. This behaviour suggests that exponential corrections do not merely shift the stability thresholds but actively restructure the thermodynamic phase space, reflecting the non-trivial nature of the quantum corrections at this level of approximation.

\begin{figure}[H]
\centering
  \begin{tabular}{@{}cccc@{}}
    \includegraphics[width=7cm]{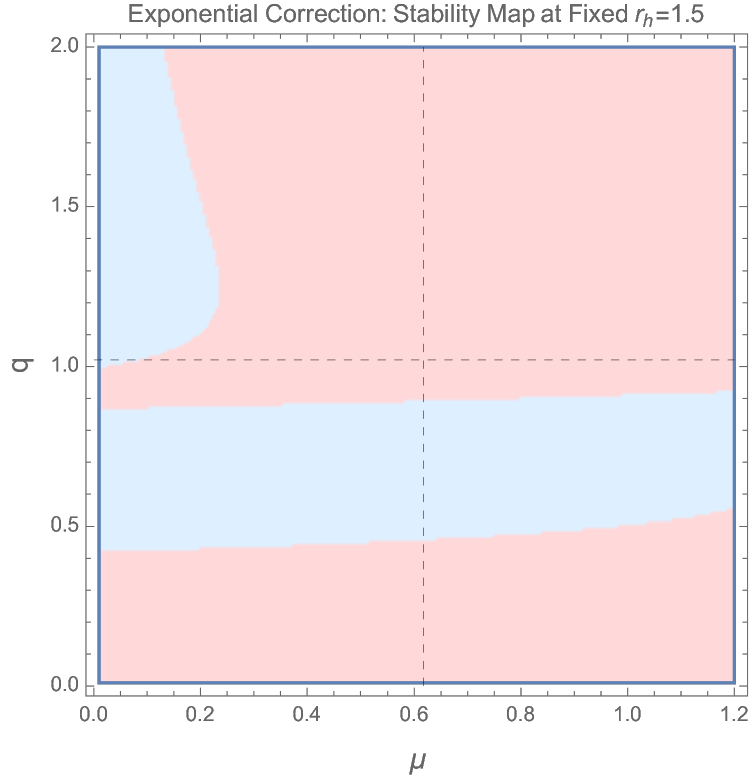} &

    \end{tabular}
 \caption{Stability map for the black hole model including exponential corrections to the entropy, displayed in the \((\mu, q)\) plane at fixed horizon radius \(r_h = 1.5\). Compared to the previous cases, the structure of the stable region (blue) is more intricate and curved, indicating strong dependence on both the NED parameter \(\mu\) and the electric charge \(q\). The red region denotes thermodynamic instability. While a portion of the stable domain lies within the bounds \(\mu = \tfrac{50}{81}\) and \(q = \sqrt{25/24}\), the phase transition boundary is no longer linear and exhibits richer structure, reflecting the substantial influence of exponential quantum corrections on the thermodynamic phase space.}\label{figgg3}
\end{figure}

\subsubsection{Physical Significance of Thermodynamic Variables in  Einstein-Euler-Heisenberg Theory  }

In black hole thermodynamics one may distinguish five complementary thermodynamic variables, each offering a different window onto the system’s equilibrium and stability. The internal energy, identified with the black hole’s total mass–energy in isolation, underpins the most fundamental (microcanonical) description. The Helmholtz free energy gauges the balance of heat and work when the hole is held at fixed size in a thermal bath, thereby signalling which configurations are favoured at constant volume. When the cosmological constant is reinterpreted as a pressure, the black hole’s mass naturally assumes the role of enthalpy, representing the energy required both to assemble the hole and to displace the surrounding spacetime. In extended black hole thermodynamics, the pressure is defined via the identification \( P = -\Lambda / 8\pi \), where \( \Lambda \) denotes the cosmological constant. This approach promotes \( \Lambda \) from a fixed geometric parameter to a thermodynamic variable, thereby extending the first law and Smarr relation to include a \( P\,\mathrm{d}V \) term. Under this prescription, the black hole mass \( M \) is naturally reinterpreted as the enthalpy of the system rather than its internal energy. This interpretation has been widely adopted and justified in the literature (see, e.g., \cite{kubizvnak2012p}).The Gibbs free energy governs processes at fixed temperature and pressure, determining whether a black hole will form or evaporate under isothermal–isobaric conditions. Finally, treating pressure as a dynamical variable in its own right exposes a rich phase structure, with the conjugate “thermodynamic volume” measuring the effective spacetime excluded by the black hole.\vspace{0.7em}

In what follows, we provide a comparative evaluation of the thermodynamic quantities investigated in this work. Classical terms provide the baseline structure governed by general relativity and NED. The logarithmic corrections account for leading-order quantum effects, enhancing stability and modifying phase transitions, while the exponential ones deliver a deeper, non-perturbative treatment essential for capturing strong-field phenomena, especially near the quantum gravity regime. Together, these formulations offer a progressively refined understanding of black hole thermodynamics in quantum-modified settings. For an Einstein-Euler-Heisenberg black hole satisfying the conditions taken into consideration in this paper, these results can be summarised as given in Table \ref{tabb}.

\begin{table}[H]
  \centering
  \footnotesize
  \caption{Comparison of thermodynamic quantities under classical, logarithmically corrected (C), and exponentially corrected (EC) formulations.}
  \label{tab:thermo-comparison}
  \begin{tabularx}{\textwidth}{@{} l >{\raggedright\arraybackslash}X >{\raggedright\arraybackslash}X >{\raggedright\arraybackslash}X @{} }
    \hline
    \toprule
    \textbf{Quantity}    & \textbf{Classical}                                                                                     & \textbf{Log. Correction (C)}                                                         & \textbf{Exp. Correction (EC)}                                                                                \\
    \midrule
    \hline
    $F$                 & Decreases with $\mu$ due to a $-\mu/r_h^5$ term, enhancing stability for small $r_h$.             & Logarithmic terms further lower $F$ and introduce mild scale dependence.               & Non-perturbative contributions  strongly suppress $F$ at small/intermediate $r_h$.    \\
    \addlinespace
    $G$                 & Slight increase with $\mu$, interpreted as enthalpy in extended phase space.                        & Scale-dependent analytic corrections from trace-anomaly effects.                     & Exponential terms dominate at large $r_h$, capturing nonlocal vacuum dynamics.                                 \\
    \addlinespace
    $P$                 & Enhanced at small $r_h$ by EH corrections; smooth classical profile at large $r_h$.                 & Logarithmic and polynomial modifications refine pressure profile.                    & Exponential enhancements at small $r_h$ indicate significant vacuum backreaction.                            \\
    \addlinespace
    $H$                 & Grows with $\mu$, corresponding to increased mass (enthalpy) in extended thermodynamics.              & Moderate log corrections yield gradual increase in $H$.                             & Highly nonlinear quantum terms cause steep growth of $H$ in high-curvature regimes.                             \\
    \addlinespace
    $E$                 & Decreases at small $r_h$ as $\mu$ increases, reflecting vacuum-induced energy reduction.              & Polynomial corrections align with semiclassical entropy shifts.                      & Non-analytic exponential and \textit{erfi} terms significantly modify energy content near quantum scales.        \\
    \bottomrule
    \hline
  \end{tabularx}\label{tabb}
\end{table}

\section{Astrophysical Applications}\label{sec4}
The theory of NED finds numerous applications in the field of astrophysics, as the universe is filled with sources of strong electromagnetic field. As examples, one can consider neutron stars and magnetars, since both hold the ability of producing magnetic field whose magnitude can reach up to $10^6-10^9$ T \cite{breton2023type}. In this regard, one can look for the observational implications of the theoretical discussions covered so far with the aid of two key astrophysical phenomena: the gravitational lensing and the gravitational redshift.

\subsection{Gravitational Lensing around Einstein-Euler-Heisenberg Black Holes}
In the standard Euclidean geometry, the inner product of two distinct vectors gives us the invariant angle in between. Rindler and Ishak extended this geometric result to curved spacetimes, developing a methodology for calculating the deflection of the relativistic light \cite{rindler2007contribution}. For the lensing formulation developed by RI, the geometry diagrammed below, which features a massive object at its center, is considered. 
\begin{figure}[H]
 \centering
\includegraphics[width=100mm,scale=1]{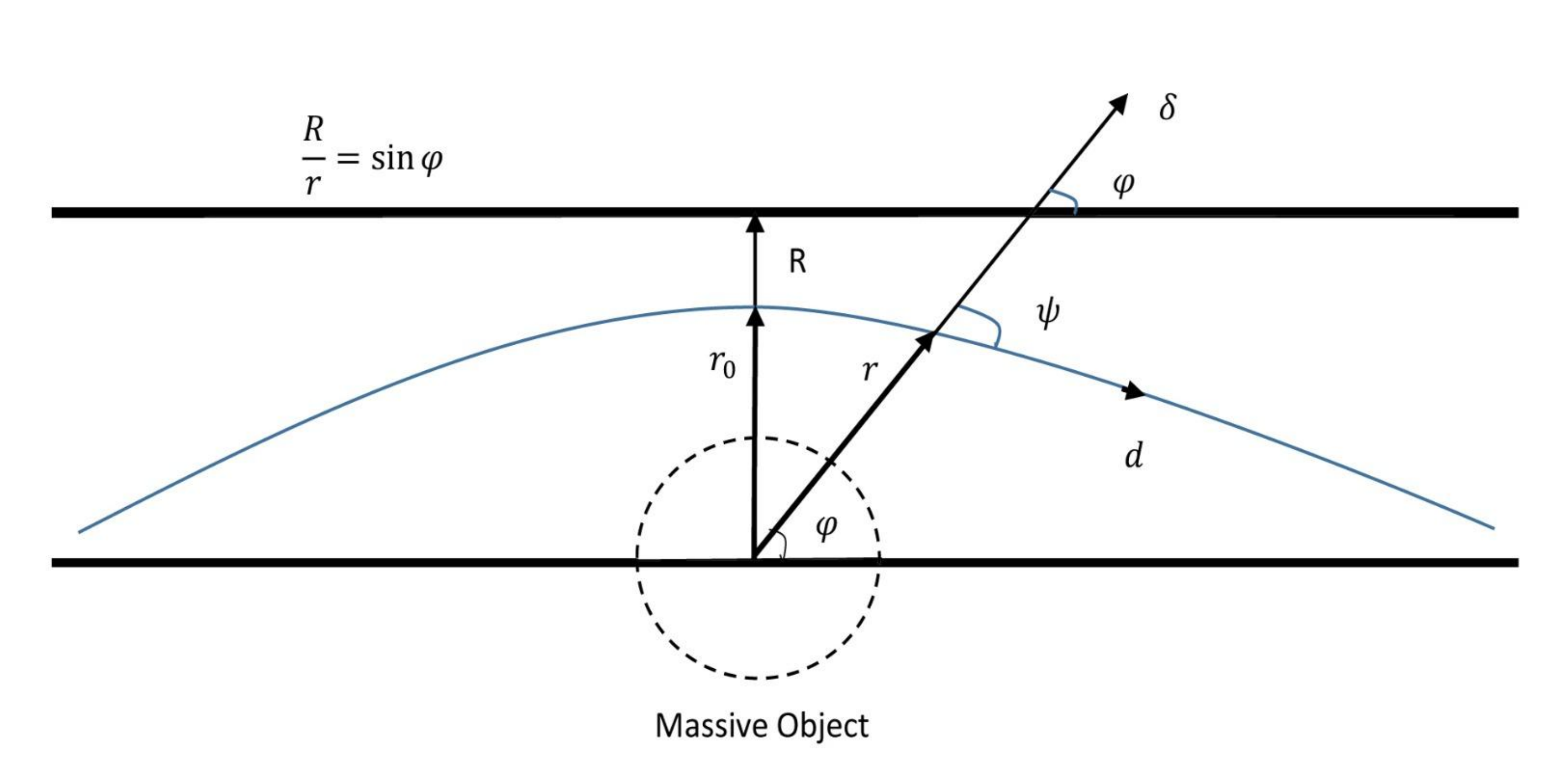}
\caption{The generic lensing structure of RI analysis}\label{figg}
\end{figure}
As seen in Fig.\ref{figg}, the angle between the two coordinate directions $\delta$ and d can be found using the invariant formula
\begin{equation}
    cos(\Psi)=\frac{d^{i} \delta_{i}}{\sqrt{(d^{i}d_{i}) (\delta^{i}\delta_{i})}}.\label{bending1}
\end{equation}
The metric tensor of the constant time slice of metric \eqref{metric1}, a two-dimensional curved (r,$\varphi$) space, is represented by $g_{ij}$ in this formula. It is expressed as the orbital plane of the light beams at the equatorial plane ($\theta=\pi/2$) in the form \eqref{57}.
\begin{equation}
    ds^{2}= \frac{dr^{2}}{f(r)} +r^{2} d \varphi ^{2}.\label{57}
\end{equation}
The formalism requires us to define the null geodesics equation. The motion constants inside the spacetime under consideration are

\begin{equation}
    \frac{dt}{d \tau} = - \frac{E}{f(r)}   \;\;,\;\;   \frac{dt}{d \varphi} = \frac{l}{r^{2}},
\end{equation}
where $\tau$, $l$ and $E$ represent the proper time, angular momentum and energy E, respectively. Subsequently, one gets
\begin{equation}
    (\frac{dr}{d\tau})^{2} = E^{2}- \frac{l}{r^{2}} f(r),
\end{equation}
and 
\begin{equation}
     (\frac{dr}{d\varphi})^{2} = \frac{r^{4}}{l^{2}} (E^{2} - \frac{l^{2}}{r^{2}} f(r)).\label{bending2}
\end{equation}
Introducing $u=1/r$, Eq. \eqref{bending2} becomes

\begin{equation}
    \frac{d^{2}u}{d\varphi^{2}}= - \frac{u^{2}}{2} \frac{d f(u)}{du} - u f(u). \label{55}
\end{equation}
At this point, based on Fig.\ref{55}, one can write

\begin{equation}
\begin{aligned}
     \delta& = (\delta r , 0) = (1,0) \delta r, \\
     d &= (dr,d\varphi) = (\frac{dr}{d\varphi},1) d\varphi.
\end{aligned}
\end{equation}
Applying these constraints to Eq.\eqref{bending1} lead to

\begin{equation}
    tan(\varphi) = \frac{[g^{rr}]^{\frac{1}{2}}r}{\vert\frac{dr}{d\varphi}\vert}.\label{58}
\end{equation}

Thus, $\varepsilon=\psi-\varphi$ can measure the one-sided bending angle of the curved spacetime. When metric \eqref{metric} is substituted in Eq. \eqref{55}, the null geodesic equation of Euler-Heisenberg spacetime takes the form

\begin{equation}
    \frac{d^{2}u}{d\varphi^{2}}+u= 3Mu^2-2Q^2u^3+\frac{\bar{\mu} Q^4}{5}u^7\label{60}
\end{equation}

Now, the solution of Eq.\eqref{60} can be analysed. Since finding the exact solution of such a nonlinear differential equation is rather challenging, the standard approximate solution method can be used to obtain a perturbative solution of Eq. \eqref{60}. In this context, substituting $u(\varphi)=sin(\varphi)/R$, the linear and homogeneous solution of Eq. \eqref{60}, into the nonlinear part on the right-hand side of Eq. \eqref{60}, the first-order approximate solution becomes

\begin{equation}
\begin{aligned}
    u = \frac{1}{r} = &\frac{\sin \varphi}{R} + \frac{M}{R^2} \left( 1 + \cos^2 \varphi \right) + \frac{3Q^2}{R^3} \left( \frac{\varphi \cos \varphi}{4} - \frac{\cos^2 \varphi \sin \varphi}{12} - \frac{\sin \varphi}{6} \right) \\
    & + \frac{\bar{\mu} Q^4}{40 R^7} \left( \sin \varphi - \frac{35\varphi}{16} - \frac{19 \cos^4 \varphi \sin \varphi}{24} + \frac{29 \cos^2 \varphi \sin \varphi}{16} \right). \label{u1}
\end{aligned}
\end{equation}
The first-order derivative of $r(\varphi)$ with respect to $\varphi$ becomes ,

\begin{equation}
\begin{aligned}
    \frac{dr}{d\varphi} = r^2 \Big[ & -\frac{\cos \varphi}{R} + \frac{Q^2}{4R^3} \left( 3\varphi \sin \varphi - \cos \varphi - \sin 2\varphi \sin \varphi + \cos^3 \varphi \right) \\
    & + \frac{M \sin 2\varphi}{R^2} + \frac{\bar{\mu} Q^4}{40R^7} \left( \frac{35}{16} - \cos \varphi - \frac{19 \cos \varphi \sin^2 2\varphi}{24} + \frac{29 \sin 2\varphi \sin \varphi}{16} \right) \Big].\label{u2}
\end{aligned}
\end{equation}
Note that in Eqs. \eqref{u1} and \eqref{u2} the integration parameter is called the impact parameter which is generally large $(R>>1)$. Additionally, taking $\varphi = \pi/2$ in Eq.\eqref{u1}, the closest approach distance is found as

\begin{equation}
    u(\varphi=\pi/2)=\frac{1}{r_0}=\frac{1}{R}+\frac{M}{R^2}- \frac{Q^2}{2R^3}+\frac{\bar{\mu} Q^4\left(32-35\pi \right)}{1280 R^7}.
\end{equation}
Now, the small angle $\psi_0$ can be calculated by setting $\varphi=0$. In this case, from Eq.\eqref{u1} and Eq.\eqref{u2}, one can write

\begin{equation}
    r(\varphi=0)\approx\frac{R^2}{2M} \;\; , \;\;  \mathcal{A}(r,\varphi=0)\approx-\frac{r^2}{R}. \label{u3}
\end{equation}

Once all the information is plugged into the Rindler-Ishak lensing formula using small angle approximation $(tan(\psi_0)\approx \psi_0)$, Eq.\eqref{58} gives

\begin{equation}
\begin{aligned}
    \varepsilon \simeq \frac{2M}{R} [g^{rr}]^{\frac{1}{2}}&\simeq \frac{2M}{R} \left( 1-\frac{4M^2}{R^2}+\frac{4M^2Q^2}{R^4}-\frac{16\bar{\mu} Q^4 M^6}{5R^{12}}\right)^\frac{1}{2} \\ & \simeq \frac{2M}{R} \left( 1-\frac{2M^2}{R^2}+\frac{2M^2Q^2}{R^4}-\frac{8\bar{\mu} Q^4 M^6}{5R^{12}}\right)+\mathcal{O}\left( \frac{M^{13}\bar{\mu}^2Q^8}{R^{25}}\right). \label{u4}
\end{aligned}
\end{equation}
    
 \subsection{Gravitational Redshift around Einstein-Euler-Heisenberg Black Holes}

Another significant observational phenomenon known as gravitational redshift, which is observable in an area with lesser gravitational field, is caused by light or other electromagnetic radiation emitted from a source in a strong gravitational field shifting to longer wavelengths. This phenomena is predicted by general relativity. The  gravitational redshift analysis in references \cite{glendenning2012compact} and \cite{zubairi2015static}  describe the spacetime geometry surrounding a non-rotating black hole. This approach indicates that the formula for the gravitational redshift can be expressed as

\begin{equation}
z=\frac{\lambda_{o}-\lambda_{e}}{\lambda_{e}}=\frac{\lambda_{o}}{\lambda_{e}}-1=\frac{\omega_{e}}{\omega_{o}}-1, \label{red1}
\end{equation}

where

\begin{equation}
\frac{\omega_{e}}{\omega_{o}}=\sqrt{g_{tt}}.  \label{red2}
\end{equation}

Here,  $\lambda_{e}$  and  $\lambda_{o}$  represent the emitted  and  the observed  wavelengths, respectively, whereas $g_{tt}$ corresponds to the $-tt$ component of the relevant metric tensor. Similarly, $\omega_{e}$ and $\omega_{o}$ denote the emitted and the observed frequencies. When we put Eq.\eqref{nmetric} into Eq.\eqref{red1}, the gravitational redshift of the Einstein-Euler-Heisenberg black Hole  becomes

\begin{equation}
    z= \frac{M}{R}-\frac{q^2}{2R^2}+\frac{\bar{\mu} Q^4}{40R^6}. \label{red3}
\end{equation}
\subsection{Astrophysical Applications: Strong Electrically Charged Compact Stars}
In the light of the information obtained, the relevant astrophysical applications are discussed. In this purpose, the numerical analysis has been established  for three realistic charged compact stars, with their properties detailed in Table \ref{tabb1}.

\setlength{\tabcolsep}{4pt}
\begin{table}[H]
  \centering
  \footnotesize
  \caption{The numerical values of the mass, radius, and charge of three compact stars. Here, $M_{\odot}$ represents the mass of the Sun \cite{gurtug2019effect}.}
  \label{tab:compact-stars}
  \begin{tabular}{@{} l @{\hskip -0.05pt} l @{\hskip -0.05pt} l @{\hskip -0.09pt} l @{} }
    \toprule  \hline
    \textbf{Compact Star $\;\;\;\;$$\;\;\;\;$$\;\;\;\;$} & \textbf{M} ($M_{\odot})\;\;\;\;$$\;\;\;\;$$\;\;\;\;$ & \textbf{R (km)$\;\;\;\;$$\;\;\;\;$$\;\;$} & \textbf{Q (C)} \\  \hline
    \addlinespace % Adds vertical space under header
    \midrule
    Vela X-1 & $1.77 M_{\odot}$ & $9.56$ & $1.81\times 10^{20}$ \\ 
    SAX J1808.4–3658$\;\;\;\;$ & $1.435 M_{\odot}$$\;\;\;\;$ & $7.07$$\;\;\;\;$ & $1.87\times 10^{20}$ \\ 
    4U 1820-30 & $2.25 M_{\odot}$ & $10$ & $1.89\times 10^{20}$ \\  \hline
    \bottomrule
  \end{tabular} \label{tabb1}
\end{table}

Note that in the numerical analyses to be performed using Eq.\eqref{u4} and Eq.\eqref{red3} obtained in standard geometric units, we should use the factors  $Gc^{-2}$ $(M)$ and $G^{1/2}c^{-2}\left( 4\pi
\varepsilon _{0}\right) ^{-1/2}$ for the mass $(M)$ and the electric charge $(q)$, respectively, based on the values provided in Table-1. Here, $G=6.67408\times10^{-11}m^{3}kg^{-1}s^{-2}$ represents the gravitational constant, $c=3\times10^{8}ms^{-1}$ denotes the speed of light and $\varepsilon _{0}=8.85418\times10^{-12}C^{2}N^{-1}m^{-2}$ is the vacuum permittivity. Note that all the resulting graphical analyses will be in SI units, and for all the cases, the boundary condition $\bar{\mu}/M^2 \leq 50/81$ will be applied as a fundamental constraint. Accordingly, all numerical evaluations will be performed by adopting the upper limit, leading to the condition $\bar{\mu} = 50M^2/81$. This approach ensures that the analyses consistently operate within the specified parameter space, facilitating a systematic investigation of the nonlinear effects while maintaining adherence to the imposed boundary constraints. Additionally, it should be noted that this choice allows for the consideration of the maximum nonlinear effects. Also note that the boundary condition derived from the solution  $Q^2/M^2 \leq 25/24$ is satisfied for the numerical values corresponding to the compact objects under consideration. This condition ensures consistency with the physical and mathematical framework applied to model such compact astrophysical structures.

Fig.\eqref{fig58} illustrates the one-sided bending angle for three electrically charged compact objects: Vela X-1, SAXJ1808.4-3658, and 4U1820-30.  As evident from Fig.\eqref{fig58} , the presence of electric charge significantly influences the bending angle of light.

\begin{figure}[H]
\centering
\begin{tabular}{@{}cccc@{}}
    \includegraphics{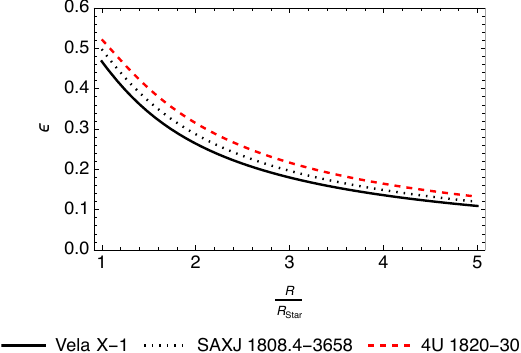} 
  \end{tabular}
  \caption{The plots of deflection angle $\varepsilon$ versus $R/R_{star}$ for each compact objects in Table \ref{tabb1}.}\label{fig58}
\end{figure}

Fig.\eqref{fig57} presents the gravitational redshifts, plotted individually for each compact object.

\begin{figure}[H]
\centering
\begin{tabular}{@{}cccc@{}}
    \includegraphics{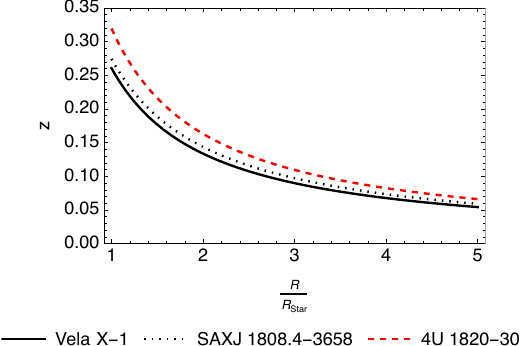} 
\end{tabular}
  \caption{The plots show the variation of the gravitational redshift with respect to $R/R_{star}$ for each compact object listed in Table  \ref{tabb1}.} \label{fig57}
\end{figure}

\section{CONCLUSION}

In this study, the nonlinear Euler-Heisenberg theory has been investigated from the perspectives of the general theory of relativity and astrophysics. We have initiated our discussions by applying logarithmic and exponential corrections to the classical Hawking-Bekenstein entropy and then checked for the consequent effects arising due to these modifications mainly via the inspection of the graphical illustrations. The concerned graphs have been obtained based on the detailed analytical evaluations of the thermodynamic parameters of the Einstein-Euler-Heisenberg black hole. The comparisons of the uncorrected and logarithmically corrected Helmholtz free energy and pressure against horizon graphs have revealed interesting information regarding the properties of the concerned black hole. Firstly, the general behaviour of the corrected functions are found to act as the reflections of the uncorrected ones through the $x$-axis. Furthermore, recalling that the logarithmic corrections arise due to small fluctuations around the equilibrium, we suggest that such corrections incorporate information regarding the interaction of the Einstein-Euler-Heisenberg black hole with vacuum. Accordingly, in Fig \ref{fig0}, the effect of vacuum fluctuations are observed to become apparent, especially for large  NED parameter values. In other words, as the nonlinear effects become dominant, making further comments on the corrections due to vacuum polarisation gets relatively easier. \\

On the other hand, the logarithmic corrections have not seem to have caused any significant change in the overall classical thermodynamic behaviour of the black hole, whereas in the exponential case, the graphs of the associated quantities have been found to behave in a completely different manner. For instance, in the positive domain of the heat capacity versus horizon graphs, the non-perturbative quantum corrections have been observed to introduce discrete vertical lines for all values of $q$. The strikingly different patterns evident in all exponentially corrected thermodynamic graphs make it tempting to suggest that the dominance of quantum effects might be the underlying reason behind the revelation of such a discrete behaviour.  In a similar context, this behaviour might suggest the forbiddance of certain states within certain horizon values, i.e. the quantisation of the thermodynamic states of the Einstein-Euler-Heisenberg black hole.\\

Based on the associated plots, we have seen that to distinguish the differences between the effect of the logarithmic and exponential corrections on the thermodynamic states of the Einstein-Euler-Heisenberg black hole, one could check the behaviour of the Helmholtz free energy versus black hole horizon for various $q$ and $\mu$ values. For $r_h>1$, the uncorrected and the logarithmically corrected Helmholtz free energy function converge for all $q$ and $\mu$ values, whereas the exponentially corrected ones diverge from one another. \\  

In addition to the investigation of the thermodynamic quantities via the Smarr formula, we have estimated the one-sided bending angle and the gravitational redshift of electromagnetic radiation in the vicinity of astronomical structures obeying the nonlinear Euler-Heisenberg model. We believe that this study sheds light on the influence of the logarithmic and the exponential corrections on entropy, which are both crucial when it comes to the determination of the complete form of entropy to be used in black hole thermodynamics and any relevant application.

 A comparative analysis of the three entropy models—classical Bekenstein-Hawking, logarithmic correction, and exponential correction—reveals a clear hierarchy in thermodynamic structure and sensitivity to parameters. In the uncorrected case, the stability landscape is simple and broad, with a well-defined horizontal stability band in the \((\mu, q)\) plane. This baseline scenario reflects the classical balance between gravitational attraction and electromagnetic repulsion and admits a wide range of stable black hole configurations within the physically admissible bounds.

 In contrast, the logarithmic correction narrows the stability region significantly, though it largely preserves the qualitative features of the classical case. The band remains horizontal and only weakly dependent on the NED parameter \(\mu\), indicating that the leading-order quantum correction primarily affects the quantitative bounds of stability without introducing additional structural complexity.

 The exponential correction, however, induces a markedly different behaviour. The resulting stability region becomes strongly dependent on both \(\mu\) and \(q\), exhibiting a curved and dynamically structured form. The associated phase transition boundary evolves from a single monotonic curve into a non-linear, potentially multi-zonal transition line, reflecting a more profound restructuring of the thermodynamic phase space. While stable configurations still exist within the physically admissible window, they are considerably more constrained and sensitive to parameter variation.

 Taken together, these findings illustrate that quantum corrections to black hole entropy can play a critical role in shaping the macroscopic thermodynamic properties of the system. The extent and nature of this influence depend strongly on the functional form of the correction, with exponential modifications introducing the most significant departures from classical behaviour.

During the astrophysical applications, the spacetime background is assumed to obey the Euler-Heisenberg model due to the presence of strong electromagnetic fields. The one-sided deflection angle of light is found to be markedly influenced by the electric charge of compact objects. The comparison presented therein considers three astrophysical systems—Vela X-1, SAX~J1808.4--3658, and 4U~1820--30—each possessing distinct mass-to-radius ratios and charge estimates. The results indicate that the presence of electric charge introduces a non-negligible enhancement in the light-bending angle, particularly for compact objects with smaller radii and larger surface charge values.

 Furthermore, the gravitational redshift corresponding to the Einstein-Euler-Heisenberg black hole has been obtained as $z = \frac{M}{R} - \frac{Q^2}{2R^2} + \frac{\bar{\mu} Q^4}{40R^6}$, which encapsulates the leading-order quantum correction to the classical result. The final term arises due to vacuum polarisation effects and reflects the influence of Euler-Heisenberg NED in the weak-field regime. The correction becomes particularly relevant in the vicinity of highly compact, charged objects whose electric field is much smaller than the Schwinger field and may offer observational signatures in future high-precision astrophysical measurements.

\bibliography{ref.bib}
\end{document}